\documentclass[aps,prmaterials,preprint,groupedaddress]{revtex4-2}

\usepackage[T1]{fontenc}
\usepackage[utf8]{inputenc}
\usepackage[english]{babel}

\usepackage[table]{xcolor}
\usepackage{chemformula}
\usepackage{xr}
\usepackage{hyperref}
\usepackage{cleveref}
\usepackage{siunitx}
\usepackage{xspace}
\usepackage{multirow}
\usepackage{hhline}
\usepackage{csquotes}
\usepackage{comment}
\usepackage{amsmath,bbm,bm,amssymb}
\usepackage{longtable}
\usepackage{lmodern}

\newcommand{\cmcm}{$Cmcm$\xspace}
\newcommand{\cmcii}{$Cmc2_1$\xspace}
\newcommand{\piic}{$P2_1/c$\xspace}
\newcommand{\abchx}{\ch{M(II)2M(III)Ch2X3}\xspace}
\newcommand{\snbchx}{\ch{Sn2M(III)Ch2X3}\xspace}
\newcommand{\pbbchx}{\ch{Pb2M(III)Ch2X3}\xspace}
\newcommand{\pimmch}{MMCH\xspace}
\newcommand{\pimmchs}{MMCHs\xspace}

\newcommand{\xs}{\ch{X}-site\xspace}
\newcommand{\chs}{\ch{Ch}-site\xspace}
\newcommand{\miii}{\ch{M(III)}\xspace}
\newcommand{\mii}{\ch{M(II)}\xspace}
\newcommand{\x}{\ch{X}\xspace}
\newcommand{\chch}{\ch{Ch}\xspace}
\newcommand{\miiis}{\ch{M(III)}-site\xspace}
\newcommand{\miis}{\ch{M(II)}-site\xspace}

\DeclareSIUnit\at{atom}
\DeclareSIUnit\angstrom{\text {Å}}
\setchemformula{kroeger-vink=true}

\begin{document}

\author{Pascal Henkel}
\affiliation{Department of Applied Physics, Aalto University, P.O.Box 11100, FI-00076 AALTO, Finland}

\author{Jingrui Li}
\affiliation{State Key Laboratory for Manufacturing Systems Engineering; Electronic Materials Research Laboratory, Key Laboratory of the Ministry of Education, School of Electronic Science and Engineering; International Joint Laboratory for Micro/Nano Manufacturing and Measurement Technology, Xi'an Jiaotong University, Xi'an 710049, China}

\author{Patrick Rinke}
\affiliation{Physics Department, Technical University of Munich, Garching, Germany}
\affiliation{Atomistic Modelling Center, Munich Data Science Institute, Technical University of Munich, Garching, Germany}
\affiliation{Munich Center for Machine Learning (MCML)}
\affiliation{Department of Applied Physics, Aalto University, P.O.Box 11100, FI-00076 AALTO, Finland}
\email{patrick.rinke@tum.de}

\title{Design Rules for Optimizing Quaternary Mixed-Metal Chalcohalides}

\date{\today}

\begin{abstract}
Quaternary mixed-metal \abchx chalcohalides are an emerging material class for photovoltaic absorbers that combines the beneficial optoelectronic properties of lead-based halide perovskites with the stability of metal chalcogenides. Inspired by the recent discovery of lead-free mixed-metal chalcohalides materials, we utilized a combination of density functional theory and machine learning to determine compositional trends and chemical design rules in the lead-free and lead-based materials spaces. We explored a total of 54 \abchx materials with \mii = \ch{Sn}, \ch{Pb}, \miii = \ch{In}, \ch{Sb}, \ch{Bi}, \chch = \ch{S}, \ch{Se}, \ch{Te}, and \x = \ch{Cl}, \ch{Br}, \ch{I} per phase (\cmcm, \cmcii, and \piic). The \piic phase is the equilibrium phase at low temperatures, followed by \cmcii and \cmcm. The fundamental band gaps in \cmcm and \cmcii are smaller than those in \piic, but direct band gaps are more common in \cmcm and \cmcii. The effective electron masses in \piic are significantly larger compared to \cmcm and \cmcii, while the effective hole masses are nearly the same across all three phases. Using random forest regression, we found that the two electron acceptor sites (\ch{Ch} and \ch{X}) are crucial in shaping the properties of mixed-metal chalcohalide compounds. Furthermore, the electron donor sites (\ch{M(II)} and \ch{M(III)}) can be used to finetune the material properties to desired applications. These design rules enable precise tailoring of mixed-metal chalcohalide compounds for a variety of applications.
\end{abstract}

\maketitle

\section{Introduction}
Photovoltaic (PV) technologies are ideal for providing clean, affordable, and secure energy, facilitating a shift towards sustainable energy production. Continuous development of new PV materials is essential to enhance power conversion efficiency (PCE), extend device longevity, and reduce costs, thereby promoting the widespread adoption of solar cells. Metal halide perovskites have emerged as promising candidates.\cite{feng2023,zhang2016,NRELchart} Specifically, lead halide perovskites (LHPs) offer cost-effective fabrication, advantageous optoelectronic properties, and high defect tolerance.\cite{seo2016,nie2020a,savill2021} However, the commercial viability of LHPs is challenged by lead toxicity and their moderate long-term stability in air.\cite{huang2021, huang2021a, ganose2017, nie2020a} Low toxicity alternatives to LHPs include \ch{Sn^{2+}}-, \ch{Sb^{3+}}- or \ch{Bi^{3+}}-based perovskites. However, these materials face challenges such as high defect densities and oxidation in air (\ch{Sn^{2+}} to \ch{Sn^{4+}}).\cite{cao2021} Metal chalcogenides based on \ch{Pb^{2+}}, \ch{Cd^{2+}}, or \ch{Sb^{3+}}\cite{im2011, lee2009, choi2014, kavanagh2021a} have emerged as promising LHP alternatives due to their high absorption cross sections and tunable band gaps. Stable solar cells have already been demonstrated for these materials, albeit with modest efficiencies.\cite{NRELchart,choi2014,im2011,yadav2023}

Perovskite-inspired quaternary mixed-metal chalcohalides (\pimmchs) that combine halide perovskite (\ch{ABX3}) and metal chalcogenide (\ch{MCh2}) building blocks are a promising semiconductor material class for photovoltaic applications.\cite{nie2020b, kavanagh2021b, sopiha2022} Their stoichiometry is \abchx (or  \ch{A2BCh2X3} in short)\cite{kavanagh2021b} where the \ch{M(II)}- and \ch{M(III)}-sites are occupied by metals with $ns^2$ lone pair electron(s), the \chs by chalcogens, and the \xs by halogens. The properties of both building blocks may combine synergistically in mixed-metal chalcohalides as evidenced by a high defect tolerance owing to the  strong dielectric screening (due to $ns^2$ lone pair electrons), the resultant low capture cross-sections of defects, and dispersive valence and conduction band edges.\cite{huang2021, huang2021a, nie2020b} Also, the strong metal-chalcogen bond, which is formed in \pimmch materials between the bivalent chalcogenide anions and both \ch{M(II)} and \ch{M(III)} metal cations,  can overcome the known stability challenges of LHPs.\cite{nie2020b, nie2020a} In addition, the strength of the \ch{Sn}-\ch{Ch} bond should prevent the oxidation from \ch{Sn^{2+}} to \ch{Sn^{4+}} in lead-free \pimmch-based compounds. As a result, \snbchx films can be stable in humid conditions when synthesized under reduction conditions, as verified by experimental X-ray photoelectron spectroscopy.\cite{nie2020b, kavanagh2021b} Recently, Nie \textit{et al.} achieved a PCE of \SI{4.04}{\percent} for a \ch{Sn2SbS2I3}-based single-junction solar cell,\cite{nie2020b} which highlights the PV potential of mixed-metal chalcohalides. \SI{4.04}{\percent} is a promising start considering that perovskite solar cells initially showed a PCE of \SI{3.8}{\percent} in 2009\cite{kojima2009} and now exceed $\SI{26}{\percent}$.\cite{NRELchart}

In this work, we focus on the \pimmch materials space and apply density functional theory (DFT) and machine learning to assess its potential for PV applications. Recently, we explored  solely the lead-free \pimmch materials with  DFT and discovered 24 new materials\cite{henkel2023} bringing the total up to 27 (including the already known \ch{Sn2SbS2I3},\cite{olivierfourcade1980, nie2020b, kavanagh2021b, nicolson2023, roth2024} \ch{Sn2SbSe2I3},\cite{ibanez1984} and \ch{Sn2BiS2I3}\cite{islam2016}) compounds. Now we extend our study to the lead-based \pbbchx materials space, reporting 25 previously unstudied compounds alongside the two known materials, \ch{Pb2SbS2I3}\cite{dolgikh1985, doussier2007, roth2023} and \ch{Pb2BiS2I3},\cite{islam2016, roth2023} which had been the only Pb-based representatives reported to date. We compute the equilibrium properties of the 27 lead-based counterparts to the tin compounds of our previous study with DFT. We then apply a random forest (RF) machine learning model and the Shapley additive explanations (SHAP) analysis to discover chemical trends for lead-free as well as lead-based perovskite-inspired quaternary mixed-metal chalcohalides.

Our goal is to go beyond a purely DFT-based analysis by leveraging a combination of RF models and SHAP analysis to identify and understand the root of chemical trends of the four atomic sites and their elements on material properties for a total of 54 lead-free and lead-based \pimmch compounds. To cover a broad range of \pimmch materials, we considered \ch{Sn^{2+}} and \ch{Pb^{2+}} (both with $ns^2$ lone pair electrons) for the \miis, and \ch{Sb^{3+}}, \ch{Bi^{3+}} (both with $ns^2$ lone pair electrons), and In(III) (with a $5d^{10}6s^0$ valence electron configuration) for the \miiis. Additionally, we included \ch{S^{2-}}, \ch{Se^{2-}}, and \ch{Te^{2-}} for the \chs, and \ch{Cl^-}, \ch{Br^-}, and \ch{I^-} for the \xs. We considered three phases (\cmcm, \cmcii, and \piic) in which \pimmch materials can crystallize to ensure structural diversity. This allows us to assess the dominance of electron acceptor sites (\ch{Ch} and \ch{X}) over electron donor sites (\ch{M(II)} and \ch{M(III)}). An additional objective is to determine whether and how the chemical trends depend on the material parameters, such as the consistent decrease in formation energies (thermodynamic stabilities), band gaps, as well as effective electron and hole masses from \ch{Cl} via \ch{Br} to \ch{I}. Finally, we will investigate the effect of the \pimmch structure and phase on the chemical trends.

The structure of this article is as follows: In Section\,\ref{sec:comp_details}, we outline our computational workflow, starting with data generation using DFT, followed by the creation of appropriate RF models to examine the influence of atomic sites, and concluding with an analysis of the elements' impact on material properties using SHAP. Section\,\ref{sec:results} reports the effects of atomic sites and elements on material properties. Section\,\ref{sec:discuss} discusses the identified effects and design rules on on hand within the \pimmch materials space and on the other hand in terms of chemical origin. Finally, Section \ref{sec:conclusion} provides a summary of our findings.

\section{Models and computational Details}\label{sec:comp_details}
We used a combination of different methods to study trends in the \pimmch materials space. Here we describe how we calculated the material properties  (formation energy, fundamental band gap, optical band gap, and effective electron and hole masses) for different \pimmchs crystal structures by means of DFT. Next, we outline how we determined the influence of atomic sites (\ch{M(II)}, \ch{M(III)}, \ch{Ch} and \ch{X}) on the materials properties with RF regression models. Finally, we present details of the SHAP analysis to assess the impact of different elemental substitutions on the materials properties.

\subsection{\protect\abchx structures}\label{sec:comp_details:abchx_struc}

\begin{figure}[ht]
\includegraphics[scale=0.28]{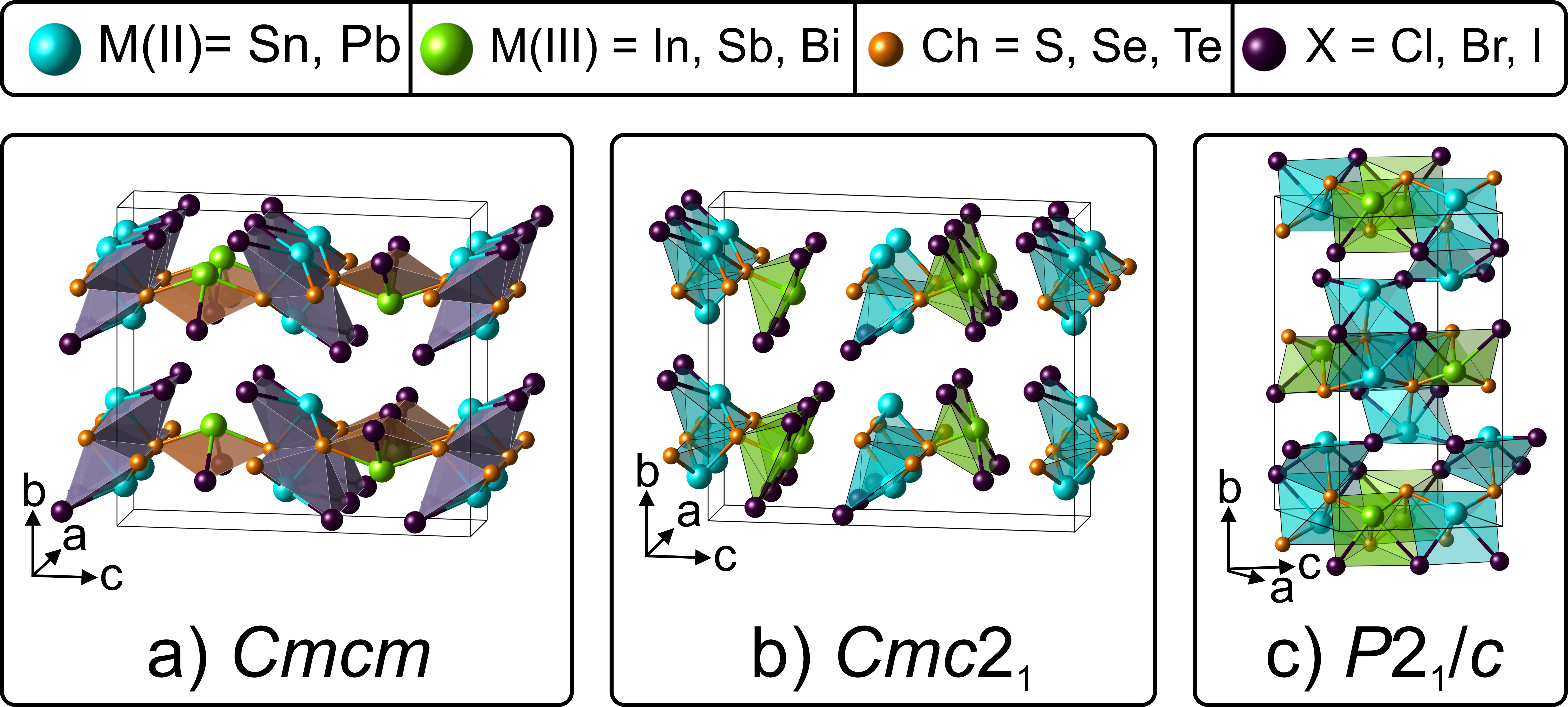}
\caption{Illustration of different phases; a) \cmcm, b) \cmcii, and c) \piic in \pimmch structures. The respective \ch{M(II)} coordination polyhedrons are depicted in cyan and the \ch{M(III)} coordination polyhedrons in green. The coordinates of each phase were taken from the Materials Project\cite{materialsproject2013} and were visualized with VESTA.\cite{momma2008}}
\label{fig:a2bch2x3_spacegrp}
\end{figure}

The structural framework of \abchx materials consists of four distinct atomic sites (see Figure\,\ref{fig:a2bch2x3_spacegrp}): the two metal sites \ch{M(II)} and \ch{M(III)}, the chalcogen site \ch{Ch}, and the halogen site \ch{X}. Both metal sites are occupied by $ns^2$ lone pair metals, except \ch{In} for \ch{M(III)}, serving as electron donors, while the \ch{Ch}- and \ch{X}-sites function as electron acceptors.

\pimmch materials can crystallize, as revealed by X-ray diffraction measurements and DFT calculations, in three different phases: \cmcm (\#{}63), \cmcii (\#{}36), and \piic (\#{}14),\cite{olivierfourcade1980, ibanez1984, dolgikh1985, doussier2007, islam2016, roth2023, roth2024, nicolson2023, kavanagh2021b, henkel2023} see Figure\,\ref{fig:a2bch2x3_spacegrp}. Throughout the manuscript, the three \pimmch structures are distinguished by their respective space group labels. The orthorhombic \cmcm phase is prevalent at room temperature,\cite{olivierfourcade1980, ibanez1984, dolgikh1985, doussier2007, islam2016, roth2023} and  transitions to a monoclinic \piic phase below \SI{100}{\kelvin}, as demonstrated by Doussier \textit{et al.} for \ch{Pb2SbS2I3}.\cite{doussier2007} DFT calculations for \ch{Sn2SbS2I3} suggest that the \cmcm structure, is thermodynamically unstable at \SI{0}{\kelvin} due to imaginary phonon modes, can be viewed as an average of the more energetically favorable, lower symmetry \cmcii phase.\cite{kavanagh2021b} Ab-initio molecular dynamic simulations further demonstrated a transition from \cmcm to \cmcii phase at \SI{500}{\kelvin},\cite{kavanagh2021b} highlighting the entropic stabilization effect of the \cmcm phase. Additionally, the \piic phase is energetically favorable over \cmcm and \cmcii.\cite{nicolson2023} A more detailed description of the structural characteristics of \pimmch materials crystallizing in the three phases (\cmcm, \cmcii and \piic) is given in Section SM\;1 in the Supplemental Material (SM).\cite{suppinfo}

The initial structures for all \pimmchs have been generated based on known materials listed in the Materials Project.\cite{materialsproject2013} For the \cmcm and \cmcii phases we used the corresponding \ch{Sn2SbS2I3} structures (ref. numbers: mp-561134 and mp-1219046) and for \piic phases the corresponding \ch{Pb2SbS2I3} structure (ref. number: mp-578882). We carried out substitutions on the \ch{M(II)}-, \ch{M(III)}-, \ch{Ch}-, and \ch{X}-sites to generate a total of 54 \pimmchs per phase.

\subsection{Density functional theory calculations}\label{sec:comp_details:dft}
We determined the material properties (formation energy, fundamental band gaps, optical gaps, effective electron and hole masses) of all 54 \pimmchs in the three phases for our RF-based analysis with DFT. We performed spin unpolarized DFT calculations with the all-electron, numeric atom-centered orbital code FHI-aims.\cite{blum2009, havu2009, ren2012, levchenko2015, yu2018, ihrig2015} We chose the  Perdew-Burke-Ernzerhof exchange-correlation functional for solids (PBEsol) \cite{perdew2008, perdew2009} due to its computational efficiency and its good agreement with experimental lattice constants for various halide perovskites\cite{yang2017, bokdam2017, laakso2022, pan2024, li2024} and the five known \pimmchs. On average, PBEsol underestimates the lattice constants for the known \pimmchs by $\sim \SI{1.7}{\percent}$ for \cmcm, $\sim \SI{1.6}{\percent}$ for \cmcm and $\sim \SI{0.4}{\percent}$ for \piic. Atomic geometries were relaxed with the Broyden-Fletcher-Goldfarb-Shanno algorithm\cite{knuth2015} in two steps. We first pre-optimized each structure with \textit{light} real-space grid settings and a tier-1 basis and then refined the geometry  using \textit{tight} settings and a tier-2 basis. The structures were relaxed until all forces acting on the atoms were smaller than \SI{5e-3}{\electronvolt\per\angstrom}. The electronic self-consistency convergence threshold was set to \SI{1e-6}{\electronvolt}. A Gaussian broadening of \SI{0.01}{\electronvolt} was applied to all electronic occupations and relativistic effects were included based on the zero-order regular approximation.\cite{blum2009, vanlenthe1993} In addition to PBEsol geometry optimization, we performed single point calculations for the band structures, the absorption spectra and the density of state (DOS) with the range-separated hybrid Heyd-Scuseria-Ernzerhof (HSE06) functional (with \SI{25}{\percent} exact exchange)\cite{heyd2003, heyd2006, krukau2006} including spin-orbit coupling\cite{huhn2017} and a tier-2 basis. The effective masses were extracted from the band structure with  a quadratic least-squares fit to the respective band edges assuming a parabolic dispersion with  the implementation of Whalley in the \textit{effmass} package\cite{whalley2018}. In addition, the absorption spectra were calculated based on the linear macroscopic dielectric tensor\cite{draxl2006} within the independent particle approximation. 

We used \num{16}-atom unit cells for the \cmcm (\#{}63) (see Figure\,\ref{fig:a2bch2x3_spacegrp}a) and \cmcii (\#{}36) (see Figure\,\ref{fig:a2bch2x3_spacegrp}b phases and a \num{32}-atom unit cell for the \piic (\#{}14) (see Figure\,\ref{fig:a2bch2x3_spacegrp}c)) phase. For PBEsol geometry optimization, we sampled the Brillouin zone with a $\Gamma$-centered $11 \times 11 \times 3$ $k$-point mesh for the \cmcm and \cmcii phases and $6 \times 3 \times 5$ for \piic. For the HSE06 band structure and absorption spectra calculations, we increased the grids to  $\Gamma$-centered $16 \times 16 \times 4$ for \cmcm and \cmcii and  $9 \times 4 \times 8$ for \piic.

The complex nature of \pimmch compounds makes a rigorous stability analysis highly demanding, as it requires knowledge of all competing phases within the quaternary phase space. As a computationally feasible DFT-based approximation, we estimate the thermodynamic stability (at \SI{0}{\kelvin}) by calculating the formation energy with respect to elemental decomposition of an individual \pimmch compound as 
\begin{equation}\label{eq:formationenergy}
E_\text{form}[\ch{M(II)2M(III)Ch2X3}] = E_\text{tot}[\ch{M(II)2M(III)Ch2X3}] - \sum_i x_i \mu_i.
\end{equation}
$E_\text{tot}[\ch{M(II)2M(III)Ch2X3}]$ is the total energy of \abchx, $x_i$ the number of atoms of the $i$th element, and $\mu_i$ the corresponding chemical potential. The upper limit for the  chemical potential is given by $\mu_i \leq E_\text{tot}(i\text{th element})$, i.e., the total energy per atom in the most stable pure phase.\cite{vandewalle2003} We used the following elemental compounds with the same computational parameters but  adjusted $\Gamma$-centered \textit{k}-point meshes: \ch{Sn} $I4_1/amd$ (\#{}141), \ch{Pb} $Fm\overline{3}m$ (\#{}225), \ch{In} $I4/mmm$ (\#{}139), \ch{Sb} $R\overline{3}m$ (\#{}166), \ch{Bi} $R\overline{3}m$ (\#{}166), \ch{S} $Fddd$ (\#{}70), \ch{Se} $P2_1/c$ (\#{}14), and \ch{Te} $P3_121$ (\#{}152), as well as \ch{Cl}, \ch{Br}, and \ch{I} in the gas phase as \ch{X_2}.

In the interest of open materials science\cite{himanen2019} we make all relevant data publicly available; see reference \citenum{nomad_snbchx} for lead-free and \citenum{nomad_pbbchx} for lead-based \abchx materials.

\subsection{Random forest regression modeling}\label{sec:comp_details:randomforest}
We used RF regression of the DFT data as a means to determine the impact of each atomic site (\ch{M(II)}, \ch{M(III)}, \ch{Ch}, and \ch{X}) on different material properties. RF is a non-linear and non-parametric ensemble learning method composed of different decision trees.\cite{breiman2001} For regression models, as used in this study, the predicted target is averaged over the individual decision trees, which increases prediction accuracy and reduces overfitting.\cite{james2013}

For each phase, we first created a data set that contains the material properties for each \pimmch material. As a descriptor we use the atomic numbers of all elements within each \pimmch, e.g., \ch{Sn2SbS2I3} is represented by the array $[50, 51, 16, 53]$. Next, we trained an RF regression model for each material property using the atomic numbers as input and the respective material property as the target value. To determine the optimal RF hyperparameters, we used the cross-validated grid-search (\textit{GridSearchCV}) method implemented in \textit{scikit-learn}.\cite{scikitlearn} A $\SI{75}{\percent}/\SI{25}{\percent}$ split for the training/test data was used to determine the performance of each hyperparameter combination. The mean absolute error (MAE) and root mean square error between predicted and DFT-calculated material properties are calculated for each hyperparameter set to assess the performance. In addition, a 10-fold cross validation was carried out to ensure the robustness of each hyperparameter set. The best performing hyperparameter set is applied to analyze the entire data set. The performance of the RF models for the formation energy, (fundamental) band gap, effective electron and hole masses for different phases is presented in Figure\,SM\;1. In addition, we also trained RF models for the lowest direct band gaps and optical band gaps, see Figures\,SM\;2 and SM\;3 for their individual performance. We used the final trained RF models to determine the importance of the atomic sites for each material property. Each atomic site is one feature of our input space and we used the built-in feature importance tools of \textit{scikit-learn}. 

Additionally, we also tested other descriptors such as electro-negativity (EN), electron affinity (EA), ionization energy (IE) as well as atomic, Van der Waals, and covalent radii. In the past, descriptors based on such tabulated properties have been successfully used to predict material properties for a variety of material systems, such as \ch{N}-based semiconductors,\cite{huang2019} \ch{ABX3},\cite{gladkikh2020} and \ch{A^'A^{''}B^'B^{''}X6} double perovskites.\cite{pilania2016} Gladkikh \textit{et al.} and Pilania \textit{et al.} used such element-derived descriptors in combination with kernel ridge regression model to predict the band gaps of \ch{ABX3}\cite{gladkikh2020} and \ch{A^'A^{''}B^'B^{''}X6} double perovskites.\cite{pilania2016} However, we found for \pimmch materials that the atomic numbers provided the best performance. Furthermore, we tested the sure independence screening and sparsifying operator (SISSO) approach as implemented in \textit{SISSO++},\cite{purcell2022} but the results were not as good as those of the RF.

\subsection{Shapley additive explanations analysis}\label{sec:comp_details:shap}
To examine the impact of elements on material properties within the \pimmch materials space, we analyzed the resulting RF models using the SHAP method.\cite{lundberg2017} SHAP, based on game theory and an additive feature attribution approach, explains machine learning outputs. Its purpose is to identify the contribution (SHAP value) of each feature (in our case of each element within a \pimmch compound) to the predicted target value. Various methods can approximate SHAP values. We here use \textit{TreeSHAP}, specifically designed for RFs \cite{lundberg2020, lundberg2018}, to assess the influence of the features (atomic numbers of elements within the four atom sites of \pimmch compounds) on the material parameters. For each RF model, a SHAP base value is first determined, representing the mean of the predicted values. Then, the contribution of each feature to the predicted target value is calculated relative to the SHAP base value. In this study, we use the SHAP implementation by Lundberg.\cite{shap2024}

In this study, we exclusively use the mean SHAP values for each element type to enhance interpretability. Since the contribution of each element varies depending on the specific \pimmch compound, minor deviations in element type contributions are observed. For instance, the SHAP contribution of \ch{Cl} to the formation energy is \SI{-75}{\milli\electronvolt} in \ch{Sn2SbS2Cl3} and \SI{-77}{\milli\electronvolt} in \ch{Sn2BiTe2Cl3}. Therefore, we report the mean SHAP contribution along with the corresponding variance, which characterizes the distribution around the mean value.

\section{Results}\label{sec:results}
We examine the impact of atomic sites and  elements on the formation energy, fundamental band gap, and effective electron and hole masses of \pimmch materials in their \cmcm phase. This analysis allows us to identify material property trends across the \pimmch materials space. The results for the \cmcii and \piic phases are shown in Figures\,SM\;4 and SM\;5 for the formation energy, Figures\,SM\;6 and SM\;7 for the fundamental band gap, and Figures\,SM\;14, SM\;16, SM\;15, and SM\;17 for the effective masses of electrons and holes. Additionally, Table\,SM\;I lists all material parameters for all 54 \pimmch materials across all phases.

\subsection{Formation energies}\label{sec:results:formenergy}
We have previously determined that lead-free \pimmch materials are stable with respect to elemental decomposition.\cite{henkel2023} Here we show that this stability also extends to the lead-based counterparts as evidenced by their negative formation energies, listed in Table\,SM\;I. Therefore, all 54 \pimmch materials are stable - independent of their phase. We observe that the \piic phase is most prevalent (i.e., has the highest absolute formation energies)  followed by \cmcii then \cmcm, see Table\,SM\;I.

\begin{figure}[h!]
\centering
\includegraphics[scale=1]{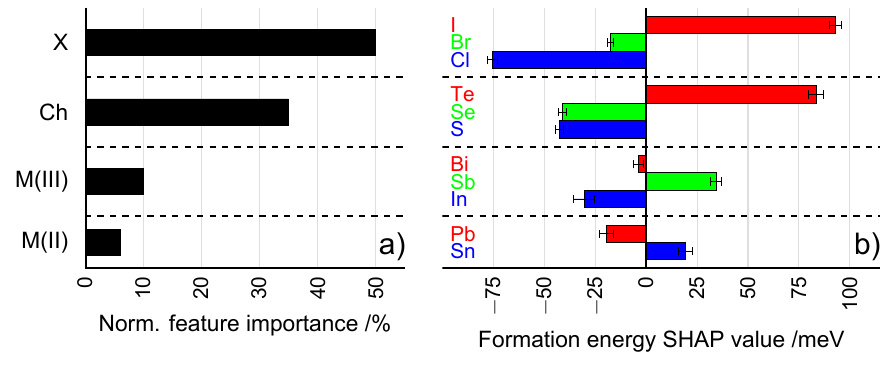}
\caption{Summary of the feature importance of the formation energy in the RF model for the \cmcm phase: a) Normalized atom site importance for the predicted formation energy in percent; b) impact of elements on the \ch{M(II)}-, \ch{M(III)}-, \ch{Ch}-, and \ch{X}-sites measured in terms of their mean SHAP values on the predicted base formation energy value in \unit{\milli\electronvolt}. The error bars denote the standard deviation for each mean SHAP value. Since the formation energy is negative, negative mean SHAP values indicate a stabilization of the material and  positive SHAP values a destabilization.}
\label{fig:results:formenergy}
\end{figure}

Figure\,\ref{fig:results:formenergy}a shows the atomic site importance of the formation energy for all 54 \pimmch materials in the \cmcm phase as determined by the RF model. The \xs has the largest influence (\SI{50}{\percent}) followed by the \chs (\SI{34}{\percent}). In contrast, both \mii (\SI{6}{\percent}) and \miii (\SI{10}{\percent}) metal sites have only minor effects on the formation energy. 

The corresponding SHAP analysis is presented in Figure\,\ref{fig:results:formenergy}b and offers more detail on the elemental contributions to the formation energy. SHAP reveals that \ch{Cl} and \ch{Br} decrease the formation energy (\ch{Cl} more than \ch{Br}) and \ch{I} increases it. On the \chs,  \ch{S} and \ch{Se} decrease the formation energy (almost by the same amount on average), while \ch{Te} increases it. For \ch{Bi}, \ch{In}, and \ch{Pb} we also observe decreases, while \ch{Sb} and \ch{Sn} increase the formation energy. In correspondence with the feature importance, the formation energy change caused by the metal atoms on \ch{M(II)}- and \ch{M(III)}-site is less than that of the \ch{X}- and \ch{Ch}-site elements. We observed a similar behavior for the  \cmcii and the \piic phases as shown correspondingly in Figures\,SM\;4 (\cmcii) and SM\;5 (\piic) in the SM.

\subsection{Fundamental band gaps}\label{sec:results:bandgaps}
\pimmch materials are mainly semiconductors, with the exception of \ch{Sn2InTe2Br3} and \ch{Sn2BiTe2I3} in the \cmcm phase (see Table\,SM\;I). The majority of the computed, fundamental band gaps are indirect. Only $\sim\SI{33}{\percent}$ of lead-free and $\sim \SI{40}{\percent}$ of lead-based \pimmch compounds feature a direct fundamental band gap in their \cmcm and \cmcii phases. Also, the vast majority of \pimmch compounds in the \piic phase have an indirect fundamental band gap. Only $\sim \SI{7}{\percent}$ of lead-free and $\sim \SI{11}{\percent}$ of lead-based \abchx compounds exhibit a direct fundamental band gap. The average values of the fundamental band gap in the \cmcm, \cmcii, and \piic phases are $\sim \num{1.2}$, $\sim \num{1.1}$, and $\sim \SI{1.6}{\electronvolt}$, respectively.

\begin{figure}[h!]
\centering
\includegraphics[scale=1]{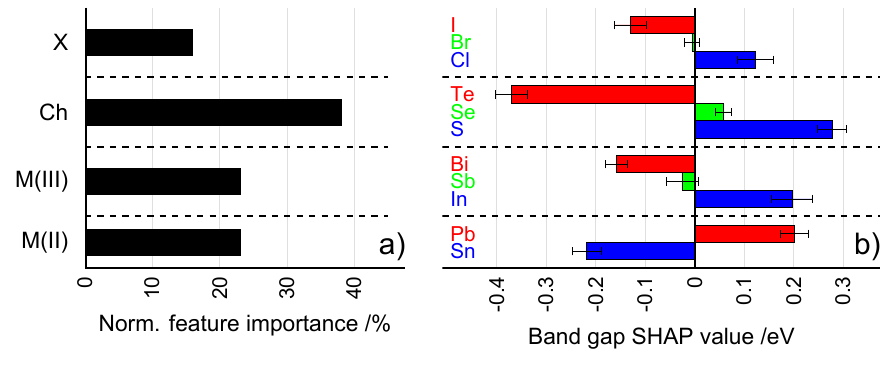}
\caption{Summary of the feature importance of the fundamental band gap in the RF model for the \cmcm phase: a) Normalized atom site importance for the predicted fundamental band gap in percent; b) impact of elements on the \ch{M(II)}-, \ch{M(III)}-, \ch{Ch}- and \ch{X}-sites quantified in terms of their mean SHAP values for the predicted base fundamental band gap in \unit{\electronvolt}.  The error bars denote the standard deviation for each mean SHAP value.}
\label{fig:results:bandgap}
\end{figure}

Figure\,\ref{fig:results:bandgap}a presents the same feature importance analysis done for the formation energy in the previous section now for the fundamental band gap in the \cmcm phase. The band gap is mainly influenced by the chalcogen site (\SI{38}{\percent}), followed by both metal sites (\SI{23}{\percent} for both \ch{M(II)} and \ch{M(III)}; see Figure\,\ref{fig:results:bandgap}a). In contrast to the formation energy, halogens have the smallest (\SI{16}{\percent}) influence on the fundamental band gap. The corresponding SHAP analysis in Figure\,\ref{fig:results:bandgap}b reveals that on the \xs \ch{Cl} and \ch{Br} decrease the fundamental band gap while \ch{I} increases it. Similarly, on the \chs \ch{S} and \ch{Se} increase the fundamental band gap and \ch{Te} decreases it.  For both metal sites, \ch{In} and \ch{Pb} lead to a band gap increase  while \ch{Bi}, \ch{Sb} and \ch{Sn} reduce the gap.

Again, we observed a similar behavior for the \cmcii and \piic phases (see Figures\,SM\;6 and SM\;7, respectively). Furthermore, we obtained similar results for an RF model trained on the lowest direct band gap for each material instead of the lowest fundamental band gap (which includes a mix of direct and indirect band gaps). The results for the lowest direct band gap are presented in Figures\,SM\;8 (\cmcm), SM\;9 (\cmcii) and SM\;10 (\piic). 

We also calculated the absorption spectra for all 54 \pimmch compounds in the three phases in the independent particle approximation. We then determined the optical band gaps as onset of absorption from the corresponding Tauc plots. In our previous work,\cite{henkel2023} we had demonstrated that the optical band gap lies on average \SI{0.29}{\electronvolt} above the fundamental band gap for the lead-free \pimmch compounds. For the lead-based systems, the difference is \SI{0.22}{\electronvolt}, which drops the average for the whole \pimmch materials system to \SI{0.26}{\electronvolt}. Figures\,SM\;11 (\cmcm), SM\;12 (\cmcii) and SM\;13 (\piic) show the corresponding feature importance analysis for the optical band gaps for all three phases. We observe that also the optical band gap is determined by the \ch{Ch}-, \ch{M(II)}- and \ch{M(III)}-sites.

\subsection{Effective electron and hole masses}\label{sec:results:effmass}

Table\,SM\;I also lists the effective electron and hole masses for the investigated \pimmch materials. Unlike for the formation energy (Section \ref{sec:results:formenergy}) and the fundamental band gap (Section \ref{sec:results:bandgaps}), the effective masses behave differently across the three different phases. For \piic, the average electron ($\num{1.42}\; m_0$ with $m_0 = \SI{9.109e-31}{\kilo\gram}$ denoting the free electron mass) and hole ($\num{1.46}\;m_0$) masses are almost the same, while for  \cmcm and \cmcii electrons are lighter than holes ($m_\text{e}^{Cmcm} = \num{0.86}\;m_0$, $m_\text{h}^{Cmcm} =  \num{1.37}\;m_0$, $m_\text{e}^{Cmc2\textsubscript{1}} = \num{0.93}\;m_0$, and $m_\text{h}^{Cmc2\textsubscript{1}} = \num{1.38}\;m_0$).

\begin{figure}[h!]
\centering
\includegraphics[scale=1]{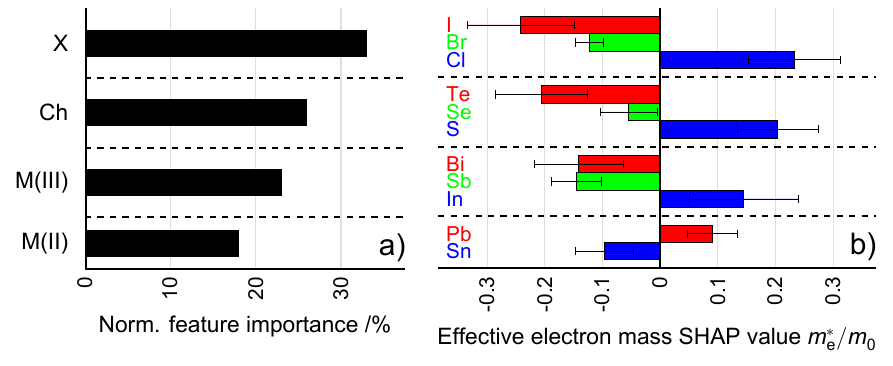}
\caption{Summary of the feature importance of the effective electron mass in the RF model for the \cmcm phase: a) Normalized atom site importance for the predicted electron mass in percent; b) impact of elements on the \ch{M(II)}-, \ch{M(III)}-, \ch{Ch}-, and \ch{X}-sites quantified in terms of their mean SHAP values for the predicted base effective hole mass value per $m_0$ (free electron mass).  The error bars denote the standard deviation for each mean SHAP value.}
\label{fig:results:effelectronmass}
\end{figure}

\begin{figure}[h!]
\centering
\includegraphics[scale=1]{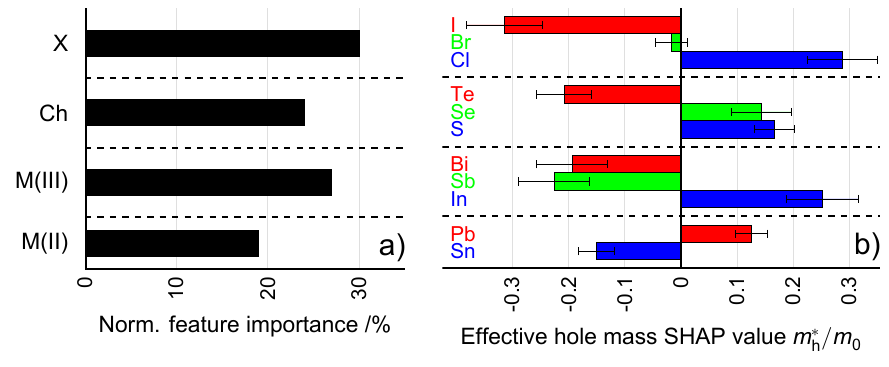}
\caption{Summary of the feature importance of the effective hole mass in the RF model for the \cmcm phase: a) Normalized atom site importance for the predicted effective hole mass in percent; b) impact of elements on the \ch{M(II)}-, \ch{M(III)}-, \ch{Ch}-, and \ch{X}-sites quantified in terms of their mean SHAP values for the predicted base effective hole mass value per $m_0$.  The error bars denote the standard deviation for each mean SHAP value.}
\label{fig:results:effholemass}
\end{figure}

The corresponding feature importance plots are shown in 
Figures\,\ref{fig:results:effelectronmass}a and \ref{fig:results:effholemass}a for \cmcm, Figures\,SM\;14a and SM\;16a for \cmcii, and SM\;15a and SM\;17a for \piic. They reveal that the \miis has the lowest impact on the effective electron and hole masses regardless of the phase. The order of importance of the other three sites varies for different phases. For the \cmcm phase, the \ch{X}-site has the largest influence, although all three sites are closer together in their effect (between \SI{23}{\percent} and \SI{33}{\percent}) than they were for the formation energy and the fundamental band gap.

Our SHAP analysis (see Figures\,\ref{fig:results:effelectronmass}b and \ref{fig:results:effholemass}b for \cmcm) reveals that in all three phases the halogens \ch{I} and \ch{Br} decrease the effective electron and hole masses whereas \ch{Cl} increase them. Similarly, on the \miis \ch{Pb} increases and \ch{Sn} decreases the effective masses. For the \cmcm phase, also \ch{Te}, \ch{Bi}, and \ch{Sb} have a decreasing effect, while \ch{S} and \ch{In} always increase both effective masses. Only \ch{Se} changes its role, providing a small decrease of the electron, but a moderate increase of the hole masses. For the other two phases, these elemental effects are slightly different and are summarized in Table\,\ref{tab:discuss:comparison} as well as Figures\,SM\;14b \&{} SM\;16b (\cmcii) and SM\;15b \&{} SM\;17b (\piic).

\section{Discussion}\label{sec:discuss}

Firstly, our analysis of the formation energy in Section~\ref{sec:results:formenergy} shows that \pimmch stability (which can be estimated from the formation energy, as more negative means more stable) is mainly influenced by the two electron acceptor sites (\ch{Ch} and \ch{X}) with the two metal sites (\ch{M(II)} and \ch{M(III)}) playing only a minor role, see Figure\,\ref{fig:results:formenergy}a. From the corresponding SHAP analysis in  Figure\,\ref{fig:results:formenergy}b we can deduce the following trends, which could be used as materials design rules: The formation energy decreases (and thus the stability increases) for halogens on the \xs along \ch{Cl} $>$ \ch{Br} $>$ \ch{I} and for chalcogens on the \chs along \ch{S} $\simeq$ \ch{Se} $>$ \ch{Te}. On the \miiis the formation energy decreases along \ch{In} $>$ \ch{Bi} $>$ \ch{Sb}. Lastly, lead-based \pimmchs are more stable than their Sn-based counterparts (\ch{Pb} $>$ \ch{Sn}), which is in good agreement with theoretical\cite{kaiser2022} and experimental\cite{pascual2020, akbulatov2019, leijtens2017} observations for \ch{Sn}- and \ch{Pb}-based perovskites.

These design rules can be rationalized by the strength of the metal-halide and the metal-chalcogen bonds. The EN of the elements, i.e., the strength with which an element can attract valence electrons, can be used as an indicator for the bond strength. Since the \ch{X} and \chs in \pimmch compounds are both electron acceptors, the bond strength increases with EN, as both sites seek to attract electron density in the bond. Therefore, for halogens (\ch{Cl} $>$ \ch{Br} $>$ \ch{I} ), the bond-strength progresses from metal-\ch{Cl} (strongest) to metal-\ch{I} (weakest) following the EN trend: \ch{Cl} (\num{3.16}) $>$ \ch{Br} (\num{2.96}) $>$ \ch{I} (\num{2.66}) on Pauling's scale.\cite{crchandbook2023} In the chalcogen family we have \ch{S} (\num{2.58}) $\simeq$ \ch{Se} (\num{2.55}) $>$ \ch{Te} (\num{2.10}), which explains the almost identical affect of \ch{S} and \ch{Se} on the formation energy and the weakest stability for the \ch{Te}-based systems. Conversely, both metal sites in \pimmchs are electron donors. With a lower EN they can transfer their valence electrons more easily to the electron acceptor sites for creating stronger chemical metal-\ch{Ch} and metal-\ch{X} bonds in the process. An inverse relation between the EN of the metal sites and stability ensues, which we see reflected in, e.g., the \miis elements \ch{In} (\num{1.78}) $<$ \ch{Bi} (\num{1.90}) $<$ \ch{Sb} (\num{2.05}). Also, \ch{Pb}-based \pimmchs are more stable than their \ch{Sn}-based analogues in line with the respective EN values (\num{1.80} and \num{1.96}). 

Figure\,\ref{fig:results:formenergy} shows that the \xs element has the largest effect on the \pimmch formation energy, followed by the \ch{Ch}-, \ch{M(III)}-, then \ch{M(II)}-sites. The difference in impact can also be rationalized by the differences in the respective EN values. For the dominant \xs they span the largest range (\num{0.50}), followed by the \ch{Ch}-, \ch{M(III)}-, then \ch{M(II)}-sites (\num{0.48}, \num{0.27}, and \num{0.16}, respectively).

Secondly, we found that the fundamental band gap (same applies to the lowest direct and optical band gaps, see Section\,\ref{sec:results:bandgaps}) are mainly influenced by the \chs and both metal sites. Furthermore, we showed that the halogens have the smallest effect on the fundamental band gap (see Figure\,\ref{fig:results:bandgap}a), a trend opposite to the formation energy. These trends can be traced back to the orbital character of the halogens, chalcogens, and both $ns^2$ lone pair metals. Halogen $p$ orbitals are generally more localized than those of the chalcogens due to their higher EN and smaller atomic size.\cite{crchandbook2023} This increased localization leads to a weaker orbital overlap and thus hybridization with the $ns^2$ lone pair orbitals of the metal sites.\cite{shi2019} In contrast, chalcogen $p$ orbitals are more diffused thus increasing the orbital overlap and the hybridization with the $ns^2$ lone pair orbitals of the metal sites, and subsequently also the impact on the valence band maximum (VBM). The chalcogens therefore have a larger effect on the fundamental band gap than the halogens. Conversely, both \ch{M(II)} and \ch{M(III)} cations have a similar orbital character ($5s^2$ lone pair and $5p$ empty orbitals for \ch{Sn} and \ch{Sb}, as well as $6s^2$ lone pair and $6p$ empty orbitals for \ch{Pb} and \ch{Bi}) except for \ch{In} (no lone pair)) and therefore their contribution to the fundamental band gap is expected to be similar. 

\begin{figure}[h!]
\centering
\includegraphics[scale=0.9]{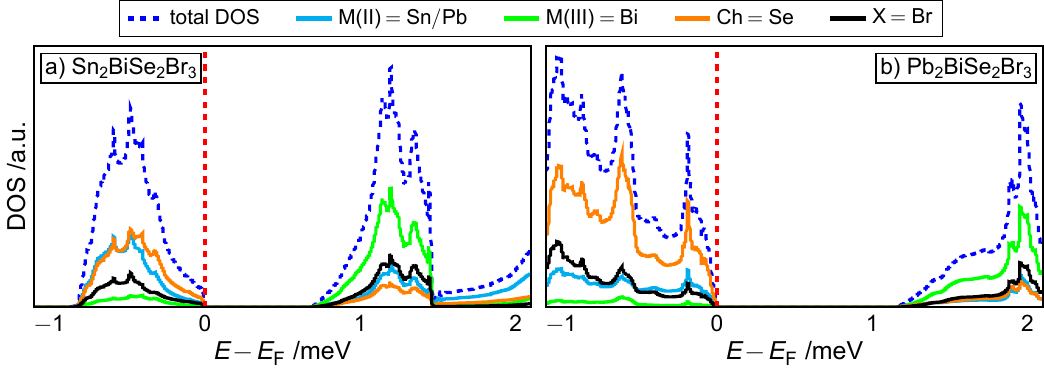}
\caption{Density of state for a) \protect\ch{Sn2BiSe2Br3} and b) \protect\ch{Pb2BiSe2Br3} at \protect\cmcm phase.}
\label{fig:dos}
\end{figure}

The effects of the orbital analysis are reflected in the DOS near the VBM and the conduction band minimum (CBM). This is illustrated in Figure\,\ref{fig:dos} for lead-based and lead-free \ch{M(II)BiSe2Br3} compounds as examples. Additionally, Figure\,SM\;18 shows the corresponding \cmcii and \piic phases. The chalcogens (orange) provide the largest contribution to the total DOS (dashed blue) near the VBM, primarily through the closed \ch{Se} $4p^6$ shell. The contribution of the \ch{M(II)} metals (cyan) is caused by the strong $ns^2$ anti-bonding character, whose important effect on the electronic structure of halide perovskites have been reported previously.\cite{frost2016, brivio2013, walsh2011} In contrast, the \ch{M(III)} metals (light green) provide the largest contribution to the total DOS near the CBM via their empty $p$ orbitals. Notably, the halogens (black) contribute significantly to the valence band  (although less than the chalcogens, primarily due to the higher EN of halogens) but only minimally to the CBM.  In summary, the hole states (near the VBM) are primarily determined by the chalcogens and the \ch{M(II)} metals, while the free electron states (near the CBM) are mainly influenced by the \ch{M(III)} metals. The fundamental band gap in \pimmchs is therefore strongly affected by the chalcogen-$ns^2$ lone pair bonds. The corresponding band structures are shown in Figure\,SM\;19. This character is different to halide perovskites for which the monovalent metal (or organic) cation provide negligible contribution to both band edges. Their impact on the electronic structure is indirect (via the crystal structure). In this regard, mixed-metal chalcohalides offer more options for tailoring the electronic structure than halide perovskites.

The design rules we derive from the SHAP analysis (see Figure\,\ref{fig:results:bandgap}b) for the band gap are generally similar to the ones for the formation energy, i.e., the larger EN for the nonmetals and the smaller EN for the metals, the higher the stability and the larger the band gap. For the \miii we observe an exception. The reverse EN trend \ch{In} $>$ \ch{Bi} $>$ \ch{Sb} for the formation energy does not hold for the band gap (\ch{In} $>$ \ch{Sb} $>$ \ch{Bi}). The absence of the $ns^2$ lone pair in \ch{In}$^{3+}$ gives rise to a larger band gap in \ch{In}-based \pimmchs, breaking the trend. Conversely, the strong spin-orbital coupling induced by \ch{Bi} significantly reduces the band gaps of \ch{Bi}-based compounds to lower than those of the isostructural \ch{Sb}-based analogs, a trend opposite to that observed for \ch{Pb} vs. \ch{Sn}. 

The observation that the fundamental band gap in \pimmch compounds decreases from \ch{Cl} via \ch{Br} to \ch{I} and from \ch{Pb} to \ch{Sn} is also found in \ch{ABX3} perovskites.\cite{hao2014, yang2016, li2019, kour2019, yang2016a} Furthermore, Nie \textit{et al.} and Islam \textit{et al.} report a decrease of the optical band gap from \ch{Sn2SbS2I3} (\SI{1.41}{\electronvolt}) to  \ch{Sn2BiS2I3} (\SI{1.22}  {\electronvolt}\cite{nie2020b,islam2016} in agreement with our observed design rule for the \miiis (\ch{Sb} $>$ \ch{Bi}). This agrees well with our calculated values, \num{1.547} and \SI{1.390}{\electronvolt}, see Table\,SM\;I. Moreover, we observe similar band gap trends in chalcogenides and sesquioxides for the \miiis. Huerta-Flores \textit{et al.} measured a band gap decrease from \ch{In2S3} to \ch{Sb2S3} to \ch{Bi2S3} (\num{2.10} $>$ \num{1.80} $>$ \SI{1.70}{\electronvolt}) \cite{huertaflores2015} that is found also in DFT calculations by Matsumoto \textit{et al.} for oxides (\ch{As2O3} > \ch{Sb2O3} > \ch{Bi2O3}).\cite{matsumoto2011} These comparisons illustrate the consistency of our band gap design rules with the available literature and across different materials spaces.

Thirdly, our results show that the conduction bands in \pimmch materials are more curved than the valence bands meaning smaller effective electron masses than hole masses, especially in the \cmcm and \cmcii phases. The curvature at the VBM is almost the same regardless of the phase. However, small changes in the chemical nature of \pimmch compounds can shift their band structure and curvature, leading to inconsistent trends in effective masses across all three phases. Despite this, we found consistent patterns for the impact of halogens and metals at the \miis on electron and hole effective masses. Specifically, the effective masses decrease in the order of \ch{Cl} $>$ \ch{Br} $>$ \ch{I} for halogens, and \ch{Pb} $>$ \ch{Sn} for metals. These trends match well with previous studies for \ch{ABX3}\cite{feng2014, chen2024, ashariastani2017} and two-dimensional halide perovskites.\cite{dyksik2020} Ashari-Astani \textit{et al.} used tight binding and $\bm{k}\cdot\bm{p}$ theory to establish a strong linear relationship between the effective electron and hole masses and the fundamental band gap in  lead-free and lead-based \ch{ABX3} perovskites. They attributed this correlation, especially for holes, to the orbital overlap between halogens on the \xs and metal atoms on the \ch{B}-site. We observe a similar correlation in \pimmchs. Regardless of the phase, effective masses decrease in the order of \ch{Cl} $>$ \ch{Br} $>$ \ch{I} for the \xs, and \ch{Pb} $>$ \ch{Sn} for the \miis, matching the fundamental band gap trends. The correlation between effective masses and band gap suggests a significant orbital contribution from the \ch{X}-$ns^2$ metals bond, consistent with Ashari-Astani \textit{et al.}'s explanation for \ch{ABX3} perovskites. However, we find no consistent correlations for the \ch{Ch}- and \ch{M(III)}-sites, which we attribute to the different chemical environments in the different phases. The increasing number of \ch{M(III)}-\ch{Ch} bonds per stoichiometric unit (one in \cmcii, two in \cmcm and three in \piic) gives rise to different coordination polyhedra and therefore a different environment around the \ch{M(III)}-\ch{Ch} bonds. Due to this difference in local environments, the effective mass trends differ across the three phases.

The fact that our purely data-driven model qualitatively agrees with the \enquote{chemical intuition} of a human scientist (e.g. based on the well-established EN or orbital character) is reassuring and illustrates that machine learning can find chemical trends without needing the auxiliary information of element specific attributes. While site-specific dependencies (e.g., \ch{Cl} $>$ \ch{Br} $>$ \ch{I} for stability and band gap) and their rationalization in terms of chemical descriptors like EN could have easily been deduced by human scientists, the quaternary nature of the \pimmch materials space makes the deduction of chemical trends more challenging. Our data-driven models revealed the chemical trends, established their relative importance (e.g. \ch{X}- and \ch{Ch}-sites dominate over the metal sites for the formation energy) and enables the formulation of a correlation between different material parameters. This allows us to deepen our \pimmch material spaces knowledge.

\begin{figure}[h!]
\centering
\includegraphics[scale=.75]{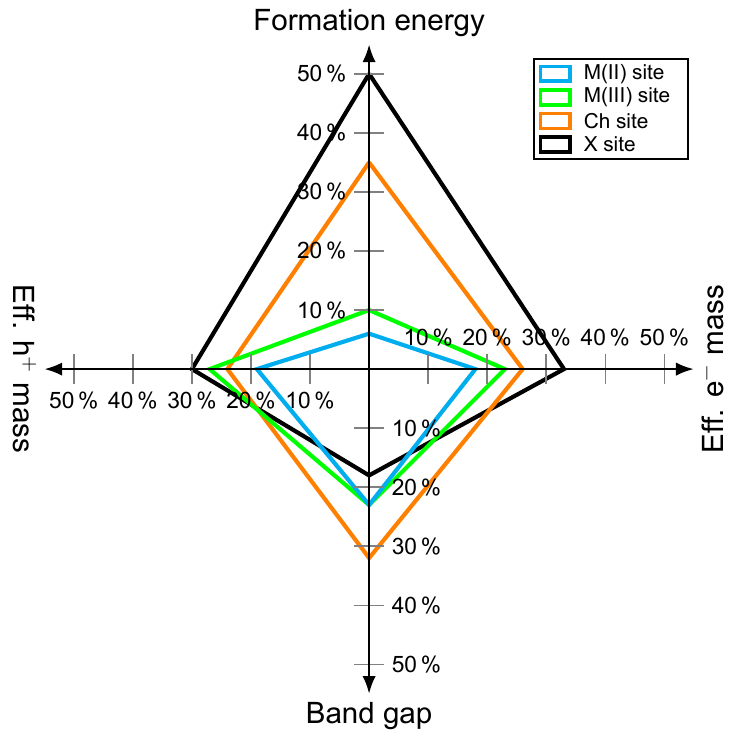}
\caption{Summary of atom site importance on the formation energy, band gap, as well as the effective masses of electrons and holes in \cmcm.}
\label{fig:discussion:vgl_featureimpoetance}
\end{figure}

Finally, we can now use the atom site importance and design rules we have derived for materials design recommendations. Figure\,\ref{fig:discussion:vgl_featureimpoetance} summaries the atom site importance's of the formation energies, fundamental band gaps, as well as of effective electron and hole masses for \cmcm. Also, see Figures\,SM\;20 (\cmcii) and SM\;21 (\piic) for the other two phases.

We have demonstrated that the two electron acceptor sites (\ch{Ch} and \ch{X}) affect the materials properties more than the two electron donor sites (\ch{M(II)} and \ch{M(III)}). Future \pimmch design studies can set the basic material properties with the two acceptor sites and finetune them for to desired applications with appropriate choices for the two donor sites. 

Furthermore it is remarkable that the elements impact across the atomic sites is fairly consistent across all  material parameters. While a correlation between band gaps and the effective masses of electrons and holes might be expected, it is rather surprising, that this relationship also extends to structural stability. 
This trend is particularly evident for choosing elements within the \ch{M(II)}- and \ch{X}-sites, see Table\,\ref{tab:discuss:comparison}. In contrast, the design rules vary to some extent at the \ch{M(III)}- and \ch{Ch}-sites (see Table\,\ref{tab:discuss:comparison}), leading to a more pronounced trade-off between stability and electronic properties. For example,  \ch{Sb} and \ch{Se} reduce the stability, but \ch{M(II)2SbSe2X3}-based material exhibit preferable band gaps for photovoltaic applications with suitable electron effective masses, yet high hole masses. In contrast, swapping \ch{Sb} for \ch{Bi} increases the stability or \ch{Se} for \ch{Te} decreases it, but both substitutions lower the electron and hole masses and reduce the band gaps. 

Overall, this allows adjustments toward either more stable materials with wider band gaps and heavier electrons and holes or less stable materials with narrower band gaps and lighter charge electrons and holes. Specifically, \ch{Pb2M(III)Ch2Cl3}-based materials tend to be more stable with wider band gaps, whereas \ch{Sn2M(III)Ch2I3}-based compounds shift to the opposite. In this context, \ch{Pb2SbSe2Cl3} and \ch{Sn2BiS2I3} stand out, combining relatively low charge carrier masses with band gaps of approximately \SI{1.4}{\electronvolt} and \SI{1.0}{\electronvolt}, respectively, making them promising candidates for further experimental and theoretical investigation.

\begin{table}[h!]
\begin{longtable}{c|cccc}
\caption{Summary of the elements' impact on the formation energy, band gap as well as effective electron and hole masses within the atom sites. Color code indicates the relative importance of the atom site. green: largest impact; olive: 2$^\text{nd}$ largest impact; yellow: 2$^\text{nd}$ smallest impact; orange: smallest impact. Deviating trends within the space groups (for the effective electron and hole masses) are indicated by 1) \cmcm, 2) \cmcii and 3) \piic, otherwise the listed element trends are identical in all three space groups.}
\label{tab:discuss:comparison}\\
& \textbf{M(II)-site} & \textbf{M(III)-site} & \textbf{Ch-site} & \textbf{X-site} \\\hline
\endfirsthead
\textbf{Formation energy} & \cellcolor{orange!100}\ch{Pb} $>$ \ch{Sn} & \cellcolor{yellow!100}\ch{In} $>$ \ch{Bi} $>$ \ch{Sb} & \cellcolor{olive!100}\ch{S} $\simeq$ \ch{Se} $>$ \ch{Te} & \cellcolor{green!100}\ch{Cl} $>$ \ch{Br} $>$ \ch{I} \\
\textbf{Band gap} & \cellcolor{yellow!100}\ch{Pb} $>$ \ch{Sn} & \cellcolor{olive!100}\ch{In} $>$ \ch{Sb} $>$ \ch{Bi} & \cellcolor{green!100}\ch{S} $>$ \ch{Se} $>$ \ch{Te} & \cellcolor{orange!100}\ch{Cl} $>$ \ch{Br} $>$ \ch{I} \\
\textbf{eff. electron mass} & \cellcolor{orange!100}\ch{Pb} $>$ \ch{Sn} & \begin{tabular}{@{}c@{}}\cellcolor{yellow!100}1) \ch{In} $>$ \ch{Bi} $\simeq$ \ch{Sb} \\ \cellcolor{yellow!100}2) \ch{In} $>$ \ch{Bi} $>$ \ch{Sb} \\ \cellcolor{yellow!100}3) \ch{Bi} $>$ \ch{Sb} $>$ \ch{In}\end{tabular} & \begin{tabular}{@{}c@{}}\cellcolor{olive!100}1) \ch{S} $>$ \ch{Se} $>$ \ch{Te} \\ \cellcolor{olive!100}2) \ch{Te} $>$ \ch{Se} $>$ \ch{S} \\ \cellcolor{olive!100}3) \ch{S} $>$ \ch{Se} $>$ \ch{Te}\end{tabular} & \cellcolor{green!100}\ch{Cl} $>$ \ch{Br} $>$ \ch{I} \\
\textbf{eff. hole mass} & \cellcolor{orange!100}\ch{Pb} $>$ \ch{Sn} & \begin{tabular}{@{}c@{}}\cellcolor{olive!100}1) \ch{In} $>$ \ch{Bi} $\simeq$ \ch{Sb} \\ \cellcolor{olive!100}2) \ch{Sb} $>$ \ch{Bi} $>$ \ch{In} \\ \cellcolor{olive!100}3) \ch{Sb} $>$ \ch{In} $\simeq$ \ch{Bi}\end{tabular} & \begin{tabular}{@{}c@{}}\cellcolor{yellow!100}1) \ch{S} $\simeq$ \ch{Se} $>$ \ch{Te} \\ \cellcolor{yellow!100}2) \ch{S} $>$ \ch{Se} $\simeq$ \ch{Te} \\ \cellcolor{yellow!100}3) \ch{Se} $>$ \ch{S} $>$ \ch{Te}\end{tabular} & \cellcolor{green!100}\ch{Cl} $>$ \ch{Br} $>$ \ch{I} \\
\end{longtable}
\end{table}

\section{Conclusion}\label{sec:conclusion}
In conclusion, this study aimed to investigate the chemical trends governing material properties in perovskite-inspired quaternary mixed-metal chalcohalides. By analyzing thermodynamic stability, band gaps, and effective electron and hole masses through a combination of DFT, RF, and SHAP, we found that the electron acceptor sites (\ch{Ch} and \ch{X}) play a primary role in determining \pimmch characteristics, while the electron donor sites (\ch{M(II)} and \ch{M(III)}) provide opportunities for fine-tuning material properties for specific applications. The choice of \ch{M(III)} and \ch{Ch} site occupants often introduces a trade-off when optimizing stability and electronic properties, whereas \ch{M(II)} and halogen sites tend to affect both properties in the same way. Our findings across three crystal structures (\cmcm, \cmcii, \piic) further highlight the importance of structural diversity, as \piic is the equilibrium phase at low temperatures with larger and mostly indirect fundamental band gaps, whereas  \cmcm and \cmcii exhibit smaller and more direct band gaps. While effective hole masses are similar in all three phases, electron effective masses are smaller in \cmcm and \cmcii compared to \piic. These results deepen our understanding of the relationship between composition, structure, and material properties in \pimmchs, opening new avenues for optimizing optoelectronic applications, such as indoor and outdoor photovoltaics. Future research could build on these identified chemical rules to explore design strategies for enhancing \pimmch performance.

\begin{acknowledgments}
The authors thank Milica Todorovi\'c, Armi Tiihonen and Jarno A. Laakso for fruitful discussions. This study was supported by the Academy of Finland through Project No. 334532, the European Union's Horizon 2020 research and innovation programme under the Marie Sk\l{}odowska-Curie grant agreement No. 101152684, the National Natural Science Foundation of China (Grant No. 22473088) and the Natural Science Foundation of Shaanxi Province of China (Grant No. 2023-YBGY-447). PH, JL and PR further acknowledge CSC-IT Center for Science, Finland, the Aalto Science-IT project, Xi'an Jiaotong University's HPC platform and the Computing Center in Xi’an of China for generous computational resources.
\end{acknowledgments}

\nocite{starosta1990}
\bibliography{lit_a2bch2x3_v2}

\end{document}


\author{Pascal Henkel}
\affiliation{Department of Applied Physics, Aalto University, P.O.Box 11100, FI-00076 AALTO, Finland}

\author{Jingrui Li}
\affiliation{State Key Laboratory for Manufacturing Systems Engineering; Electronic Materials Research Laboratory, Key Laboratory of the Ministry of Education, School of Electronic Science and Engineering; International Joint Laboratory for Micro/Nano Manufacturing and Measurement Technology, Xi'an Jiaotong University, Xi'an 710049, China}

\author{Patrick Rinke}
\affiliation{Physics Department, Technical University of Munich, Garching, Germany}
\affiliation{Atomistic Modelling Center, Munich Data Science Institute, Technical University of Munich, Garching, Germany}
\affiliation{Munich Center for Machine Learning (MCML)}
\affiliation{Department of Applied Physics, Aalto University, P.O.Box 11100, FI-00076 AALTO, Finland}
\email{patrick.rinke@tum.de}

\title{Supplemental Material for \enquote{Design Rules for Optimizing Quaternary Mixed-Metal Chalcohalides}}

\maketitle

\newpage
\pdfbookmark[section]{Description of the structural features of \protect\abchx materials in \protect\cmcm, \protect\cmcii and \protect\piic phases}{s1}
\section*{SM1: Description of the structural features of \protect\abchx materials in \protect\cmcm, \protect\cmcii and \protect\piic phases}\label{sec:si:s1}

The \pimmch \cmcm phase is characterized by the presence of \ch{M(II)Ch3X2} pyramids, which are edge-linked to form periodically continuous 1D $[\ch{M(II)2Ch2X2}]_n$ double chains along the \textit{a} axis, see main text Figure\,1a. These chains are joined via \ch{M(III)X} units and form 2D \abchx structures within the \textit{ac} plane which are not linked along the \textit{b} axis. The \cmcii structure is related to the \cmcm structure except that the \ch{M(III)X} units are linked exclusively to a single $[\ch{M(II)2Ch2X2}]_n$ chain (in contrast to \cmcm in which they are linked to 2 chains), interrupting the connection along the \textit{a} axis and preventing the formation of 2D \abchx units, see main text Figure\,1b. Similar to \cmcm, \abchx units are not connected along the \textit{b} axis in \cmcii. In the \cmcii phase, the asymmetric positioning of \ch{M(III)} between the chains causes every second square pyramidal \ch{M(II)Ch2X3} polyhedron to transform into a \ch{M(II)X3} tetrahedron thus forming a 1D $[\ch{M(II)2Ch2X}]_n$ chain, while simultaneously forming another square pyramidal \ch{M(III)ChX4} polyhedron via the attachment of \ch{M(II)X2} unit. The \piic phase differs from the other two. The primary structural features in the \piic phase are the \ch{M(II)4Ch4X8} units, which consist of four pyramids: \ch{M(II)ChX4}, \ch{M(II)Ch3X2}, \ch{M(II)Ch3X2}, and \ch{M(II)ChX4}. In these units, the two \ch{M(II)Ch3X2} pyramids share edges, while the \ch{M(II)ChX4} and \ch{M(II)Ch3X2} pyramids share faces, see main text  Figure\,1c). The neighboring \ch{M(II)4Ch4X8} units share corners (\ch{X}) within the \textit{bc} planes, and are linked via two \ch{M(III)} atoms along the \textit{c} axis to form square pyramidal \ch{M(III)Ch3X2} polyhedrons. This finally results in a 2D \ch{M(II)4M(III)2Ch4X6} structure within the \textit{bc} plane.

\newpage
\pdfbookmark[section]{Performance of the estabilised random forest regression models to predict formation energies, band gaps, as well as effective electron and hole masses}{s2}
\section*{SM2: Performance of the estabilised random forest regression models to predict formation energies, fundamental band gaps, as well as effective electron and hole masses}\label{sec:si:s2}

\begin{figure}[h!]
\centering
\includegraphics[scale=.8]{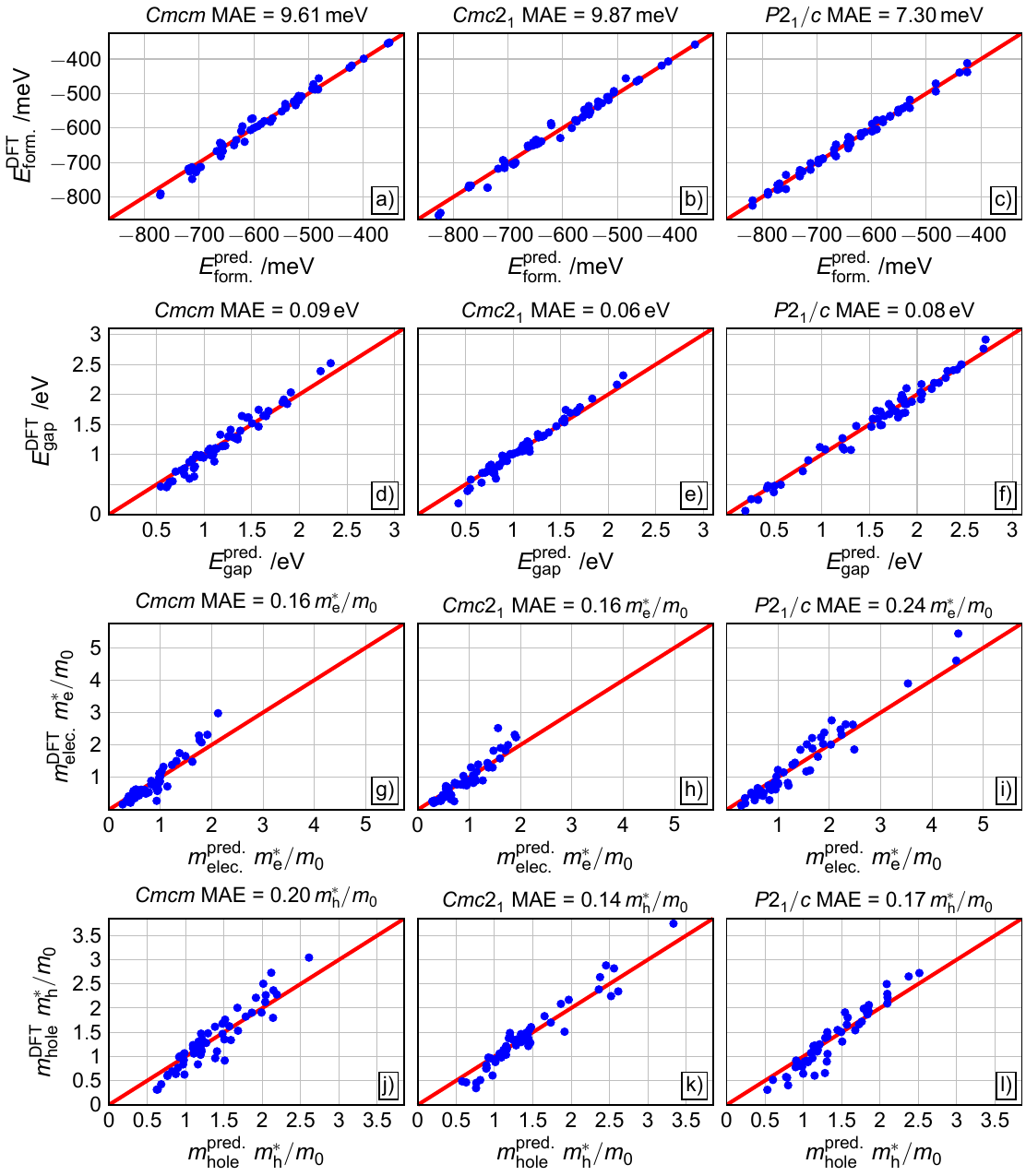}
\caption{Machine learning (random forest regression) predicted a)-c) formation energies, d)-f) fundamental band gaps, f)-i) effective electron masses and j)-l) effective hole masses compared to their DFT optimized counterparts, within the a), f) g) \&{} j) \protect\cmcm, b), e), h) \&{} k) \protect\cmcii as well as c), f), i) \&{} l) \protect\piic phase.}
\label{si:fig:rf_performance}
\end{figure}

\clearpage

\pdfbookmark[section]{Performance of the estabilised random forest regression models to predict lowest direct band gaps}{s3}
\section*{SM3: Performance of the estabilised random forest regression models to predict lowest direct band gaps}\label{sec:si:s3}

\begin{figure}[h!]
\centering
\includegraphics[scale=0.9]{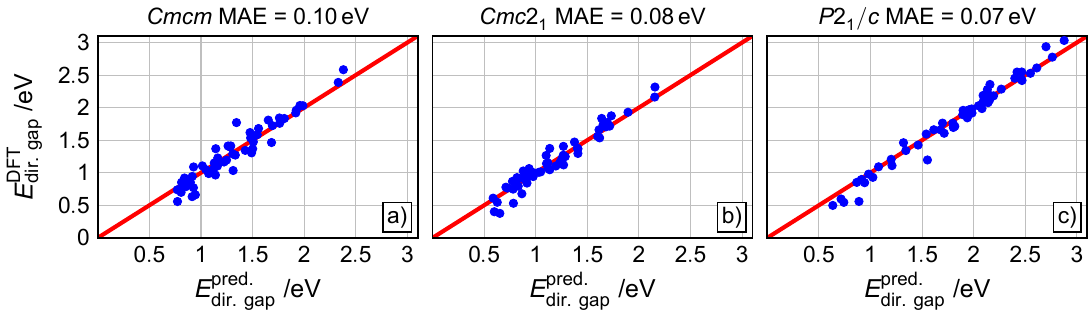}
\caption{Machine learning (random forest regression) predicted lowest direct band gaps compared to DFT optimized lowest direct band gaps within a) \protect\cmcm, b) \protect\cmcii as well as \protect\piic phase. Corresponding interpretation of results are shown in SI Figures \ref{si:fig:bandgap_cmcm:direct} for \protect\cmcm, \ref{si:fig:bandgap_cmc21:direct} for \protect\cmcii and \ref{si:fig:bandgap_p21c:direct} for \protect\piic.}
\label{si:fig:rf_performance:directbg}
\end{figure}

\pdfbookmark[section]{Performance of the estabilised random forest regression models to predict optical band gaps}{s4}
\section*{SM4: Performance of the estabilised random forest regression models to predict optical band gaps}\label{sec:si:s4}

\begin{figure}[h!]
\centering
\includegraphics[scale=0.9]{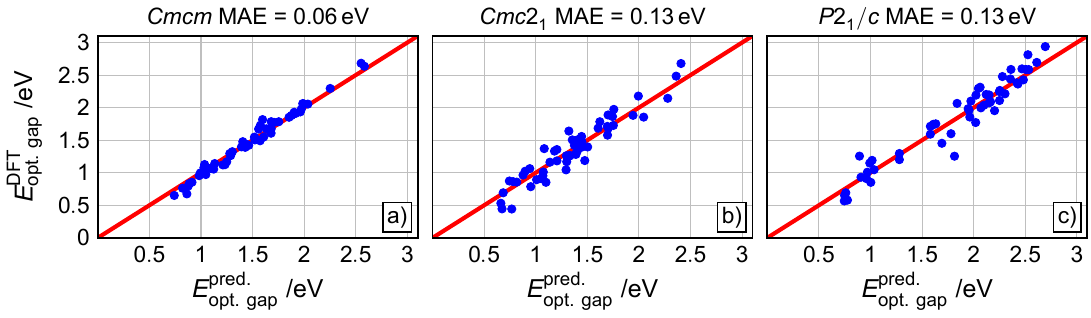}
\caption{Machine learning (random forest regression) predicted optical band gaps compared to DFT optimized optical band gaps within a) \protect\cmcm, b) \protect\cmcii as well as \protect\piic phase. Corresponding interpretation of results are shown in SI Figures \ref{si:fig:bandgap_cmcm:optical} for \protect\cmcm, \ref{si:fig:bandgap_cmc21:optical} for \protect\cmcii and \ref{si:fig:bandgap_p21c:optical} for \protect\piic.}
\label{si:fig:rf_performance:opticalbg}
\end{figure}

\clearpage

\pdfbookmark[section]{Lattice parameters, formation energies, fundamental and optical band gaps, as well as effective electron and hole masses for \protect\abchx materials}{s5}
\section*{SM5: Lattice parameters, formation energies, fundamental and optical band gaps, as well as effective electron and hole masses for \protect\abchx materials}\label{sec:si:s5}

\begin{table}[h!]
\centering
\caption{Density function theory calculated lattice parameters (\textit{a}, \textit{b}, \textit{c}) formation energies $E_\text{form}$ (related to the $i$-th elements) using PBEsol level of theory and fundamental band gaps $E_\text{g}$, and for indirect \pimmch materials also the lowest indirect band gaps ($E_\text{g, ind.}$), optical band gaps as well as effective electron and hole masses using HSE06+SOC level of theory. -/- marks that either the material has no indirect/no band gap at all or that no experimental values are available.}
\label{si:tab:matparameters}
\scalebox{.8275}{
\begin{tabular}{c|c|ccc|c|cc|c|cc}
& & \multicolumn{3}{c|}{\textbf{lat. parameters}} & & & & & \multicolumn{2}{c}{\textbf{eff. masses}}\\
\textbf{System} & \textbf{Group} & \textit{a} /\si{\angstrom} & \textit{b} /\si{\angstrom} & \textit{c} /\si{\angstrom} & $E_\text{form}$ /\si{\electronvolt\per\at} & $E_\text{g, dir.}$ /\si{\electronvolt} & $E_\text{g, ind.}$ /\si{\electronvolt} & $E_\text{g, opt.}$ /\si{\electronvolt} & $m_\text{e}^* / m_0$ & $m_\text{h}^* / m_0$ \\\hline
 & \cmcm & \num{4.23} & \num{14.12} & \num{14.30} & \num{-726} & \num{1.31} & \num{1.25} & \num{1.49} & \num{0.62} & \num{1.91} \\
 \ch{Sn2InS2Cl3} & \cmcii & \num{3.94} & \num{14.22} & \num{15.01} & \num{-772} & \num{1.83} & \num{1.69} & \num{1.89} & \num{1.99} & \num{1.38} \\
 & \piic & \num{6.27} & \num{16.19} & \num{7.88} & \num{-779} & \num{2.43} & \num{2.04} & \num{2.44} & \num{1.89} & \num{0.89} \\\hline
 & \cmcm & \num{4.22} & \num{14.8} & \num{14.50} & \num{-650} & \num{1.62} & -/- & \num{1.98} & \num{0.87} & \num{1.35} \\
 \ch{Sn2InS2Br3} & \cmcii & \num{4.11} & \num{15.16} & \num{15.53} & \num{-693} & \num{1.74} & -/- & \num{1.78} & \num{1.82} & \num{2.09} \\
 & \piic & \num{6.36} & \num{16.55} & \num{8.07} & \num{-690} & \num{1.99} & \num{1.62} & \num{2.00} & \num{0.83} & \num{1.51} \\\hline
 & \cmcm & \num{4.37} & \num{15.68} & \num{14.83} & \num{-520} & \num{1.40} & -/- & \num{1.93} & \num{2.29} & \num{2.01} \\
 \ch{Sn2InS2I3} & \cmcii & \num{4.06} & \num{14.13} & \num{19.25} & \num{-519} & \num{1.12} & \num{0.60} & \num{1.19} & \num{0.67} & \num{2.17} \\
 & \piic & \num{7.66} & \num{16.56} & \num{8.41} & \num{-565} & \num{2.28} & \num{2.10} & \num{2.31} & \num{0.52} & \num{1.91} \\\hline
 & \cmcm & \num{4.13} & \num{14.94} & \num{14.90} & \num{-718} & \num{0.66} & \num{0.60} & \num{0.68} & \num{0.35} & \num{1.82} \\
 \ch{Sn2InSe2Cl3} & \cmcii & \num{4.08} & \num{14.09} & \num{16.02} & \num{-767} & \num{1.05} & -/- & \num{1.68} & \num{1.58} & \num{1.35} \\
 & \piic & \num{6.35} & \num{16.79} & \num{8.09} & \num{-765} & \num{2.08} & \num{1.93} & \num{2.09} & \num{0.74} & \num{1.87} \\\hline
 & \cmcm & \num{4.42} & \num{14.54} & \num{15.01} & \num{-651} & \num{1.11} & -/- & \num{1.48} & \num{0.54} & \num{1.62} \\
 \ch{Sn2InSe2Br3} & \cmcii & \num{3.98} & \num{15.22} & \num{15.58} & \num{-699} & \num{1.37} & \num{1.34} & \num{1.40} & \num{0.88} & \num{0.75} \\
 & \piic & \num{6.41} & \num{17.09} & \num{8.23} & \num{-688} & \num{1.84} & \num{1.49} & \num{1.86} & \num{1.02} & \num{0.60} \\\hline
 & \cmcm & \num{4.39} & \num{15.73} & \num{15.31} & \num{-534} & \num{1.37} & \num{1.30} & \num{1.40} & \num{2.08} & \num{1.03} \\
 \ch{Sn2InSe2I3} & \cmcii & \num{4.32} & \num{15.93} & \num{16.32} & \num{-577} & \num{0.68} & -/- & \num{1.28} & \num{0.35} & \num{0.46} \\
 & \piic & \num{7.41} & \num{17.62} & \num{8.50} & \num{-567} & \num{1.95} & \num{1.79} & \num{1.96} & \num{0.50} & \num{0.79} \\\hline
 & \cmcm & \num{4.19} & \num{12.25} & \num{18.40} & \num{-542} & \num{0.56} & \num{0.46} & \num{0.65} & \num{0.21} & \num{2.21} \\
 \ch{Sn2InTe2Cl3} & \cmcii & \num{4.16} & \num{13.24} & \num{17.21} & \num{-639} & \num{0.92} & \num{0.80} & \num{1.01} & \num{2.23} & \num{1.21} \\
 & \piic & \num{6.53} & \num{17.82} & \num{8.58} & \num{-610} & \num{1.59} & -/- & \num{1.71} & \num{1.45} & \num{1.50} \\\hline
 & \cmcm & \num{4.28} & \num{12.34} & \num{18.41} & \num{-488} & -/- & -/- & \num{0.65} & -/- & -/- \\
 \ch{Sn2InTe2Br3} & \cmcii & \num{4.31} & \num{14.70} & \num{16.90} & \num{-559} & \num{0.53} & -/- & \num{0.79} & \num{0.83} & \num{0.96} \\
 & \piic & \num{6.48} & \num{18.04} & \num{8.60} & \num{-543} & \num{1.46} & -/- & \num{1.59} & \num{1.39} & \num{1.25} \\\hline
 & \cmcm & \num{4.56} & \num{15.6} & \num{16.26} & \num{-425} & \num{1.09} & \num{0.99} & \num{1.13} & \num{1.50} & \num{0.64} \\
 \ch{Sn2InTe2I3} & \cmcii & \num{4.30} & \num{15.34} & \num{16.92} & \num{-460} & \num{0.87} & \num{0.80} & \num{0.96} & \num{0.79} & \num{0.49} \\
 & \piic & \num{6.58} & \num{18.87} & \num{8.74} & \num{-438} & \num{1.11} & \num{1.08} & \num{1.20} & \num{1.23} & \num{0.91} \\\hline
 & \cmcm & \num{4.08} & \num{12.63} & \num{15.88} & \num{-649} & \num{1.68} & \num{1.61} & \num{1.77} & \num{1.03} & \num{1.24} \\
 \ch{Sn2SbS2Cl3} & \cmcii & \num{4.01} & \num{13.81} & \num{16.02} & \num{-655} & \num{1.24} & \num{1.06} & \num{1.29} & \num{0.39} & \num{2.82} \\
 & \piic & \num{6.96} & \num{16.05} & \num{8.08} & \num{-702} & \num{2.45} & \num{2.17} & \num{2.48} & \num{2.63} & \num{0.83} \\\hline
 & \cmcm & \num{4.10} & \num{13.47} & \num{15.40} & \num{-574} & \num{1.47} & \num{0.88} & \num{1.53} & \num{0.49} & \num{1.15} \\
 \ch{Sn2SbS2Br3} & \cmcii & \num{4.10} & \num{13.92} & \num{16.12} & \num{-592} & \num{1.25} & \num{1.06} & \num{1.34} & \num{0.28} & \num{1.33} \\
 & \piic & \num{7.01} & \num{16.06} & \num{8.19} & \num{-646} & \num{2.19} & \num{1.99} & \num{2.29} & \num{2.31} & \num{0.78} \\\hline
 & \cmcm & \num{4.24} & \num{13.84} & \num{15.76} & \num{-486} & \num{1.00} & \num{0.95} & \num{1.55} & \num{0.28} & \num{1.11} \\
 \ch{Sn2SbS2I3} & \cmcii & \num{4.26} & \num{14.22} & \num{16.24} & \num{-497} & \num{1.01} & -/- & \num{1.51} & \num{0.26} & \num{1.23} \\
 & \piic & \num{7.25} & \num{16.24} & \num{8.47} & \num{-555} & \num{1.76} & \num{1.69} & \num{1.77} & \num{0.60} & \num{0.58} \\
 & Exp.\textsuperscript{\hyperref[fn:tab_a_bottom]{\textit{a}}\protect\phantomsection\label{fn:tab_a_top}} & \num{4.25} & \num{13.99} & \num{16.38} & -/- & -/- & -/- & \num{1.41} & -/- & -/- \\\hline
 & \cmcm & \num{4.16} & \num{12.48} & \num{16.92} & \num{-646} & \num{1.41} & -/- & \num{1.82} & \num{0.78} & \num{1.06} \\
 \ch{Sn2SbSe2Cl3} & \cmcii & \num{4.07} & \num{14.01} & \num{16.45} & \num{-651} & \num{1.00} & -/- & \num{1.26} & \num{1.39} & \num{1.51} \\
 & \piic & \num{7.06} & \num{17.98} & \num{8.06} & \num{-692} & \num{2.18} & \num{1.84} & \num{2.20} & \num{2.38} & \num{2.06} \\\hline
\end{tabular}}
\end{table}

\clearpage

\begin{table}[h!]
\centering
\scalebox{.8275}{
\begin{tabular}{c|c|ccc|c|cc|c|cc}
& & \multicolumn{3}{c|}{\textbf{lat. parameters}} & & & & & \multicolumn{2}{c}{\textbf{eff. masses}}\\
\textbf{System} & \textbf{Group} & \textit{a} /\si{\angstrom} & \textit{b} /\si{\angstrom} & \textit{c} /\si{\angstrom} & $E_\text{form}$ /\si{\electronvolt\per\at} & $E_\text{g, dir.}$ /\si{\electronvolt} & $E_\text{g, ind.}$ /\si{\electronvolt} & $E_\text{g, opt.}$ /\si{\electronvolt} & $m_\text{e}^* / m_0$ & $m_\text{h}^* / m_0$ \\\hline
 & \cmcm & \num{4.13} & \num{12.94} & \num{17.75} & \num{-572} & \num{1.41} & \num{1.33} & \num{1.50} & \num{1.32} & \num{1.68} \\
 \ch{Sn2SbSe2Br3} & \cmcii & \num{4.16} & \num{14.22} & \num{16.73} & \num{-586} & \num{0.94} & \num{0.89} & \num{1.04} & \num{0.29} & \num{1.50} \\
 & \piic & \num{7.10} & \num{17.16} & \num{8.29} & \num{-631} & \num{2.03} & \num{1.64} & \num{2.10} & \num{1.64} & \num{2.03} \\\hline
 & \cmcm & \num{4.28} & \num{13.95} & \num{16.44} & \num{-473} & \num{1.06} & \num{0.80} & \num{1.06} & \num{0.40} & \num{1.61} \\
 \ch{Sn2SbSe2I3} & \cmcii & \num{4.31} & \num{14.50} & \num{16.91} & \num{-493} & \num{0.96} & -/- & \num{1.64} & \num{0.26} & \num{0.98} \\
 & \piic & \num{7.28} & \num{17.08} & \num{8.57} & \num{-547} & \num{1.89} & \num{1.67} & \num{1.98} & \num{0.82} & \num{1.31} \\
 & Exp.\textsuperscript{\hyperref[fn:tab_b_bottom]{\textit{b}}\protect\phantomsection\label{fn:tab_b_top}} & \num{4.30} & \num{14.09} & \num{17.22} & -/- & -/- & -/- & -/- & -/- & -/- \\\hline
 & \cmcm & \num{4.34} & \num{12.40} & \num{17.68} & \num{-530} & \num{0.94} & \num{0.87} & \num{1.00} & \num{1.02} & \num{0.32} \\
 \ch{Sn2SbTe2Cl3} & \cmcii & \num{4.34} & \num{12.39} & \num{17.70} & \num{-523} & \num{0.95} & -/- & \num{1.37} & \num{1.44} & \num{2.88} \\
 & \piic & \num{6.45} & \num{18.10} & \num{8.36} & \num{-588} & \num{0.56} & \num{0.48} & \num{0.65} & \num{0.47} & \num{1.99} \\\hline
 & \cmcm & \num{4.24} & \num{12.51} & \num{19.69} & \num{-456} & \num{0.70} & \num{0.49} & \num{0.76} & \num{0.43} & \num{0.69} \\
 \ch{Sn2SbTe2Br3} & \cmcii & \num{4.27} & \num{14.66} & \num{17.39} & \num{-456} & \num{1.03} & \num{0.98} & \num{1.06} & \num{0.78} & \num{1.45} \\
 & \piic & \num{6.59} & \num{18.50} & \num{8.42} & \num{-518} & \num{0.55} & \num{0.25} & \num{0.57} & \num{0.67} & \num{1.13} \\\hline
 & \cmcm & \num{4.34} & \num{13.57} & \num{19.47} & \num{-352} & \num{0.79} & \num{0.47} & \num{0.80} & \num{0.58} & \num{0.99} \\
 \ch{Sn2SbTe2I3} & \cmcii & \num{4.28} & \num{14.31} & \num{18.73} & \num{-358} & \num{0.38} & \num{0.19} & \num{0.44} & \num{0.56} & \num{0.34} \\
 & \piic & \num{7.02} & \num{18.82} & \num{8.67} & \num{-412} & \num{0.85} & \num{0.02} & \num{1.26} & \num{0.52} & \num{1.10} \\\hline
 & \cmcm & \num{3.99} & \num{13.91} & \num{16.02} & \num{-714} & \num{1.54} & \num{0.99} & \num{1.55} & \num{0.61} & \num{0.63} \\
 \ch{Sn2BiS2Cl3} & \cmcii & \num{4.07} & \num{13.72} & \num{15.98} & \num{-716} & \num{1.15} & \num{1.03} & \num{1.17} & \num{0.26} & \num{2.25} \\
 & \piic & \num{6.97} & \num{15.93} & \num{8.10} & \num{-732} & \num{2.19} & \num{1.84} & \num{2.19} & \num{4.60} & \num{1.80} \\\hline
 & \cmcm & \num{4.11} & \num{13.42} & \num{15.68} & \num{-650} & \num{1.34} & \num{0.96} & \num{1.44} & \num{0.59} & \num{1.07} \\
 \ch{Sn2BiS2Br3} & \cmcii & \num{4.11} & \num{13.42} & \num{15.68} & \num{-651} & \num{1.41} & \num{1.12} & \num{1.50} & \num{0.27} & \num{1.37} \\
 & \piic & \num{7.03} & \num{16.04} & \num{8.23} & \num{-675} & \num{1.97} & \num{1.72} & \num{2.07} & \num{3.90} & \num{1.41} \\\hline
 & \cmcm & \num{4.26} & \num{13.92} & \num{15.93} & \num{-552} & \num{0.99} & -/- & \num{1.39} & \num{0.41} & \num{1.47} \\
 \ch{Sn2BiS2I3} & \cmcii & \num{4.26} & \num{13.92} & \num{15.94} & \num{-553} & \num{1.01} & -/- & \num{1.42} & \num{0.21} & \num{1.46} \\
 & \piic & \num{7.30} & \num{16.23} & \num{8.52} & \num{-583} & \num{1.61} & \num{1.49} & \num{1.60} & \num{0.66} & \num{0.56} \\
 & Exp.\textsuperscript{\hyperref[fn:tab_c_bottom]{\textit{c}}\protect\phantomsection\label{fn:tab_c_top}} & \num{4.29} & \num{14.12} & \num{16.41} & -/- & -/- & -/- & \num{1.22} & -/- & -/- \\\hline
 & \cmcm & \num{4.03} & \num{12.5} & \num{17.07} & \num{-713} & \num{1.10} & \num{1.10} & \num{1.12} & \num{0.53} & \num{1.18} \\
 \ch{Sn2BiSe2Cl3} & \cmcii & \num{4.13} & \num{13.98} & \num{16.46} & \num{-708} & \num{0.90} & \num{0.71} & \num{0.95} & \num{2.32} & \num{1.33} \\
 & \piic & \num{6.41} & \num{17.23} & \num{8.22} & \num{-731} & \num{1.20} & \num{1.08} & \num{1.25} & \num{0.3} & \num{1.72} \\\hline
 & \cmcm & \num{4.13} & \num{13.02} & \num{17.29} & \num{-635} & \num{0.97} & \num{0.67} & \num{1.06} & \num{0.44} & \num{0.42} \\
 \ch{Sn2BiSe2Br3} & \cmcii & \num{4.21} & \num{14.14} & \num{16.74} & \num{-642} & \num{0.84} & \num{0.83} & \num{0.85} & \num{1.31} & \num{1.15} \\
 & \piic & \num{6.9} & \num{17.19} & \num{8.33} & \num{-662} & \num{1.43} & \num{1.12} & \num{1.45} & \num{0.74} & \num{1.93} \\\hline
 & \cmcm & \num{4.31} & \num{13.99} & \num{16.62} & \num{-538} & \num{1.05} & \num{0.77} & \num{1.14} & \num{0.26} & \num{0.31} \\
 \ch{Sn2BiSe2I3} & \cmcii & \num{4.35} & \num{14.47} & \num{16.87} & \num{-538} & \num{1.06} & \num{0.88} & \num{1.16} & \num{0.64} & \num{0.94} \\
 & \piic & \num{7.33} & \num{17.08} & \num{8.60} & \num{-575} & \num{1.66} & \num{1.47} & \num{1.75} & \num{2.23} & \num{1.94} \\\hline
 & \cmcm & \num{4.15} & \num{12.37} & \num{17.83} & \num{-580} & \num{0.74} & -/- & \num{1.08} & \num{1.75} & \num{1.31} \\
 \ch{Sn2BiTe2Cl3} & \cmcii & \num{4.2} & \num{12.37} & \num{17.84} & \num{-579} & \num{0.76} & \num{0.70} & \num{0.86} & \num{0.90} & \num{0.45} \\
 & \piic & \num{6.53} & \num{18.02} & \num{8.48} & \num{-612} & \num{0.50} & \num{0.48} & \num{0.58} & \num{0.36} & \num{0.64} \\\hline
 & \cmcm & \num{4.24} & \num{12.63} & \num{18.26} & \num{-510} & \num{0.85} & \num{0.72} & \num{0.85} & \num{1.19} & \num{1.01} \\
 \ch{Sn2BiTe2Br3} & \cmcii & \num{4.24} & \num{12.63} & \num{18.26} & \num{-510} & \num{0.85} & \num{0.72} & \num{0.86} & \num{0.90} & \num{1.01} \\
 & \piic & \num{6.68} & \num{18.39} & \num{8.53} & \num{-542} & \num{0.60} & \num{0.44} & \num{0.69} & \num{0.29} & \num{1.23} \\\hline
 & \cmcm & \num{4.38} & \num{13.30} & \num{18.73} & \num{-399} & -/- & -/- & \num{0.77} & -/- & -/- \\
 \ch{Sn2BiTe2I3} & \cmcii & \num{4.44} & \num{14.90} & \num{17.96} & \num{-407} & \num{0.61} & \num{0.58} & \num{0.69} & \num{0.32} & \num{1.49} \\
 & \piic & \num{7.12} & \num{18.67} & \num{8.75} & \num{-439} & \num{0.89} & \num{0.26} & \num{0.93} & \num{0.19} & \num{1.55} \\\hline
 & \cmcm & \num{4.06} & \num{15.04} & \num{14.60} & \num{-790} & \num{2.58} & \num{2.52} & \num{2.68} & \num{2.98} & \num{2.13} \\
 \ch{Pb2InS2Cl3} & \cmcii & \num{4.10} & \num{15.02} & \num{15.44} & \num{-853} & \num{2.32} & -/- & \num{2.68} & \num{1.83} & \num{1.22} \\
 & \piic & \num{6.98} & \num{16.05} & \num{8.18} & \num{-810} & \num{3.03} & \num{2.91} & \num{2.94} & \num{1.21} & \num{2.65} \\\hline
 & \cmcm & \num{4.17} & \num{15.31} & \num{14.73} & \num{-714} & \num{2.39} & -/- & \num{2.64} & \num{1.48} & \num{1.76} \\
 \ch{Pb2InS2Br3} & \cmcii & \num{4.20} & \num{15.83} & \num{15.52} & \num{-772} & \num{2.16} & -/- & \num{2.48} & \num{0.89} & \num{1.28} \\
 & \piic & \num{7.36} & \num{16.03} & \num{8.36} & \num{-736} & \num{2.94} & \num{2.76} & \num{2.81} & \num{1.39} & \num{2.50} \\\hline
 & \cmcm & \num{4.38} & \num{16.02} & \num{15.03} & \num{-588} & \num{1.84} & -/- & \num{2.29} & \num{0.27} & \num{1.48} \\
 \ch{Pb2InS2I3} & \cmcii & \num{4.35} & \num{16.90} & \num{15.47} & \num{-644} & \num{1.93} & -/- & \num{2.14} & \num{0.53} & \num{1.83} \\
 & \piic & \num{7.74} & \num{16.41} & \num{8.61} & \num{-626} & \num{2.53} & \num{2.50} & \num{2.59} & \num{2.02} & \num{1.05} \\\hline

\end{tabular}}
\end{table}

\clearpage

\begin{table}[h!]
\centering
\scalebox{.8275}{
\begin{tabular}{c|c|ccc|c|cc|c|cc}
& & \multicolumn{3}{c|}{\textbf{lat. parameters}} & & & & & \multicolumn{2}{c}{\textbf{eff. masses}}\\
\textbf{System} & \textbf{Group} & \textit{a} /\si{\angstrom} & \textit{b} /\si{\angstrom} & \textit{c} /\si{\angstrom} & $E_\text{form}$ /\si{\electronvolt\per\at} & $E_\text{g, dir.}$ /\si{\electronvolt}& $E_\text{g, ind.}$ /\si{\electronvolt} & $E_\text{g, opt.}$ /\si{\electronvolt} & $m_\text{e}^* / m_0$ & $m_\text{h}^* / m_0$ \\\hline
 & \cmcm & \num{4.05} & \num{15.38} & \num{15.20} & \num{-795} & \num{2.03} & \num{1.91} & \num{2.06} & \num{0.72} & \num{1.53} \\
 \ch{Pb2InSe2Cl3} & \cmcii & \num{4.16} & \num{14.66} & \num{16.02} & \num{-846} & \num{1.77} & \num{1.73} & \num{1.85} & \num{1.30} & \num{1.03} \\
 & \piic & \num{6.47} & \num{16.78} & \num{8.42} & \num{-825} & \num{2.55} & \num{2.39} & \num{2.59} & \num{1.85} & \num{2.72} \\\hline
 & \cmcm & \num{4.16} & \num{15.59} & \num{15.31} & \num{-715} & \num{1.96} & \num{1.87} & \num{2.05} & \num{1.65} & \num{0.84} \\
 \ch{Pb2InSe2Br3} & \cmcii & \num{4.25} & \num{15.61} & \num{16.12} & \num{-767} & \num{1.71} & -/- & \num{2.18} & \num{0.85} & \num{1.14} \\
 & \piic & \num{7.47} & \num{16.45} & \num{8.48} & \num{-777} & \num{2.09} & \num{2.01} & \num{2.11} & \num{0.79} & \num{1.66} \\\hline
 & \cmcm & \num{4.36} & \num{16.1} & \num{15.51} & \num{-606} & \num{1.72} & -/- & \num{1.93} & \num{2.31} & \num{0.76} \\
 \ch{Pb2InSe2I3} & \cmcii & \num{4.40} & \num{16.62} & \num{16.25} & \num{-647} & \num{1.56} & -/- & \num{1.88} & \num{0.45} & \num{1.38} \\
 & \piic & \num{7.89} & \num{16.67} & \num{8.79} & \num{-659} & \num{1.70} & -/- & \num{2.20} & \num{0.61} & \num{1.66} \\\hline
 & \cmcm & \num{4.27} & \num{15.46} & \num{16.15} & \num{-661} & \num{1.03} & \num{1.00} & \num{0.97} & \num{0.62} & \num{2.27} \\
 \ch{Pb2InTe2Cl3} & \cmcii & \num{4.23} & \num{13.25} & \num{17.42} & \num{-718} & \num{1.14} & \num{1.08} & \num{1.18} & \num{1.3} & \num{0.51} \\
 & \piic & \num{8.61} & \num{14.61} & \num{8.54} & \num{-726} & \num{1.66} & \num{1.59} & \num{1.74} & \num{1.15} & \num{0.40} \\\hline
 & \cmcm & \num{4.32} & \num{15.72} & \num{16.30} & \num{-600} & \num{1.15} & \num{1.05} & \num{1.07} & \num{0.59} & \num{1.28} \\
 \ch{Pb2InTe2Br3} & \cmcii & \num{4.28} & \num{14.28} & \num{17.33} & \num{-643} & \num{1.27} & \num{1.22} & \num{1.36} & \num{1.91} & \num{0.77} \\
 & \piic & \num{7.41} & \num{17.54} & \num{8.46} & \num{-644} & \num{1.76} & \num{1.73} & \num{1.74} & \num{0.75} & \num{0.31} \\\hline
 & \cmcm & \num{4.42} & \num{16.38} & \num{16.46} & \num{-488} & \num{1.23} & \num{1.06} & \num{1.32} & \num{0.58} & \num{1.24} \\
 \ch{Pb2InTe2I3} & \cmcii & \num{4.38} & \num{15.23} & \num{17.26} & \num{-528} & \num{1.15} & -/- & \num{1.33} & \num{1.06} & \num{1.42} \\
 & \piic & \num{7.78} & \num{17.77} & \num{8.78} & \num{-536} & \num{1.34} & \num{1.27} & \num{1.30} & \num{0.44} & \num{0.51} \\\hline
 & \cmcm & \num{5.37} & \num{13.9} & \num{13.52} & \num{-667} & \num{2.03} & -/- & \num{2.06} & \num{2.14} & \num{1.91} \\
 \ch{Pb2SbS2Cl3} & \cmcii & \num{4.17} & \num{13.88} & \num{16.66} & \num{-704} & \num{1.87} & \num{1.79} & \num{1.97} & \num{0.77} & \num{2.34} \\
 & \piic & \num{6.70} & \num{15.76} & \num{8.29} & \num{-776} & \num{2.78} & \num{2.50} & \num{2.69} & \num{3.76} & \num{4.47} \\\hline
 & \cmcm & \num{4.18} & \num{14.51} & \num{16.25} & \num{-609} & \num{1.83} & \num{1.65} & \num{1.88} & \num{1.38} & \num{1.11} \\
 \ch{Pb2SbS2Br3} & \cmcii & \num{4.18} & \num{14.51} & \num{16.26} & \num{-634} & \num{1.54} & -/- & \num{1.73} & \num{0.41} & \num{1.39} \\
 & \piic & \num{7.03} & \num{15.92} & \num{8.38} & \num{-721} & \num{2.55} & \num{2.41} & \num{2.60} & \num{1.89} & \num{3.96} \\\hline
 & \cmcm & \num{4.32} & \num{14.06} & \num{15.82} & \num{-517} & \num{1.27} & -/- & \num{1.78} & \num{0.37} & \num{0.62} \\
 \ch{Pb2SbS2I3} & \cmcii & \num{4.32} & \num{14.88} & \num{16.44} & \num{-536} & \num{1.37} & -/- & \num{1.86} & \num{0.25} & \num{1.06} \\
 & \piic & \num{7.36} & \num{16.30} & \num{8.59} & \num{-624} & \num{2.12} & \num{2.09} & \num{2.21} & \num{0.63} & \num{1.14} \\
 & Exp.\textsuperscript{\hyperref[fn:tab_d_bottom]{\textit{d}}\protect\phantomsection\label{fn:tab_d_top}} & \num{4.33} & \num{14.18} & \num{16.56} & -/- & -/- & -/- & \num{2.00} & -/- & -/- \\
 & Exp.\textsuperscript{\hyperref[fn:tab_e_bottom]{\textit{e}}\protect\phantomsection\label{fn:tab_e_top}} & \num{7.36} & \num{16.47} & \num{8.59} & -/- & -/- & -/- & -/- & -/- & -/- \\\hline
 & \cmcm & \num{5.81} & \num{14.29} & \num{13.36} & \num{-643} & \num{1.46} & -/- & \num{1.61} & \num{0.69} & \num{2.37} \\
 \ch{Pb2SbSe2Cl3} & \cmcii & \num{4.24} & \num{13.42} & \num{17.33} & \num{-706} & \num{1.38} & \num{1.31} & \num{1.47} & \num{0.75} & \num{2.64} \\
 & \piic & \num{7.15} & \num{18.13} & \num{8.16} & \num{-761} & \num{2.42} & \num{2.19} & \num{2.43} & \num{1.18} & \num{2.10} \\\hline
 & \cmcm & \num{4.23} & \num{14.29} & \num{17.5} & \num{-595} & \num{1.58} & \num{1.28} & \num{1.67} & \num{0.79} & \num{2.73} \\
 \ch{Pb2SbSe2Br3} & \cmcii & \num{4.24} & \num{14.43} & \num{17.49} & \num{-636} & \num{1.47} & \num{1.28} & \num{1.56} & \num{0.83} & \num{1.30} \\
 & \piic & \num{7.17} & \num{16.91} & \num{8.44} & \num{-702} & \num{2.17} & \num{1.88} & \num{2.25} & \num{2.03} & \num{2.22} \\\hline
 & \cmcm & \num{4.36} & \num{14.15} & \num{16.44} & \num{-507} & \num{1.19} & \num{1.14} & \num{1.30} & \num{0.46} & \num{1.36} \\
 \ch{Pb2SbSe2I3} & \cmcii & \num{4.35} & \num{15.44} & \num{17.56} & \num{-547} & \num{1.10} & -/- & \num{1.40} & \num{0.73} & \num{0.88} \\
 & \piic & \num{7.37} & \num{16.99} & \num{8.73} & \num{-615} & \num{1.98} & \num{1.92} & \num{2.07} & \num{2.76} & \num{2.29} \\\hline
 & \cmcm & \num{5.21} & \num{13.8} & \num{15.60} & \num{-595} & \num{1.77} & \num{1.64} & \num{1.73} & \num{0.53} & \num{0.92} \\
 \ch{Pb2SbTe2Cl3} & \cmcii & \num{4.36} & \num{12.98} & \num{18.71} & \num{-569} & \num{0.75} & -/- & \num{0.91} & \num{1.17} & \num{2.39} \\
 & \piic & \num{6.36} & \num{18.48} & \num{8.37} & \num{-648} & \num{0.93} & \num{0.90} & \num{1.01} & \num{0.46} & \num{1.18} \\\hline
 & \cmcm & \num{4.55} & \num{12.69} & \num{17.93} & \num{-508} & \num{0.77} & -/- & \num{1.67} & \num{0.25} & \num{0.96} \\
 \ch{Pb2SbTe2Br3} & \cmcii & \num{4.26} & \num{13.24} & \num{19.08} & \num{-510} & \num{0.79} & \num{0.70} & \num{0.89} & \num{0.89} & \num{1.61} \\
 & \piic & \num{6.59} & \num{18.76} & \num{8.55} & \num{-577} & \num{0.85} & \num{0.72} & \num{0.91} & \num{0.46} & \num{0.92} \\\hline
 & \cmcm & \num{4.38} & \num{13.79} & \num{19.31} & \num{-355} & \num{0.85} & \num{0.55} & \num{0.96} & \num{0.16} & \num{0.93} \\
 \ch{Pb2SbTe2I3} & \cmcii & \num{4.35} & \num{14.33} & \num{19.03} & \num{-419} & \num{0.93} & \num{0.83} & \num{1.02} & \num{0.77} & \num{1.34} \\
 & \piic & \num{7.22} & \num{18.75} & \num{8.83} & \num{-471} & \num{0.90} & \num{0.37} & \num{0.85} & \num{0.13} & \num{0.86} \\\hline
 & \cmcm & \num{4.11} & \num{12.83} & \num{15.79} & \num{-748} & \num{1.92} & \num{1.64} & \num{1.91} & \num{0.76} & \num{3.04} \\
 \ch{Pb2BiS2Cl3} & \cmcii & \num{4.18} & \num{13.77} & \num{16.68} & \num{-773} & \num{1.72} & \num{1.60} & \num{1.68} & \num{1.39} & \num{3.75} \\
 & \piic & \num{6.97} & \num{15.81} & \num{8.30} & \num{-793} & \num{2.61} & \num{2.40} & \num{2.58} & \num{5.44} & \num{1.54} \\\hline
 & \cmcm & \num{4.22} & \num{13.53} & \num{15.63} & \num{-682} & \num{1.76} & \num{1.52} & \num{1.85} & \num{0.61} & \num{1.47} \\
 \ch{Pb2BiS2Br3} & \cmcii & \num{4.25} & \num{14.66} & \num{16.92} & \num{-700} & \num{1.66} & \num{1.47} & \num{1.56} & \num{1.04} & \num{1.55} \\
 & \piic & \num{7.11} & \num{16.02} & \num{8.41} & \num{-740} & \num{2.284} & \num{2.28} & \num{2.38} & \num{1.86} & \num{0.97} \\\hline

\end{tabular}}
\end{table}

\clearpage

\begin{table}[h!]
\begin{minipage}{\textwidth}
\centering
\scalebox{.8275}{
\begin{tabular}{c|c|ccc|c|cc|c|cc}
& & \multicolumn{3}{c|}{\textbf{lat. parameters}} & & & & & \multicolumn{2}{c}{\textbf{eff. masses}}\\
\textbf{System} & \textbf{Group} & \textit{a} /\si{\angstrom} & \textit{b} /\si{\angstrom} & \textit{c} /\si{\angstrom} & $E_\text{form}$ /\si{\electronvolt\per\at} & $E_\text{g, dir.}$ /\si{\electronvolt} & $E_\text{g, ind.}$ /\si{\electronvolt} & $E_\text{g, opt.}$ /\si{\electronvolt} & $m_\text{e}^* / m_0$ & $m_\text{h}^* / m_0$ \\\hline
 & \cmcm & \num{4.33} & \num{14.1} & \num{16.03} & \num{-582} & \num{1.29} & -/- & \num{1.68} & \num{0.44} & \num{0.60} \\
 \ch{Pb2BiS2I3} & \cmcii & \num{4.33} & \num{14.12} & \num{16.05} & \num{-581} & \num{1.30} & -/- & \num{1.71} & \num{0.45} & \num{0.61} \\
 & \piic & \num{7.43} & \num{16.32} & \num{8.63} & \num{-645} & \num{1.92} & -/- & \num{2.26} & \num{0.65} & \num{1.07} \\
 & Exp.\textsuperscript{\hyperref[fn:tab_f_bottom]{\textit{f}}\protect\phantomsection\label{fn:tab_f_top}} & \num{4.32} & \num{14.26} & \num{15.49} & -/- & -/- & -/- & \num{1.60} & -/- & -/- \\\hline
 & \cmcm & \num{4.21} & \num{13.16} & \num{16.88} & \num{-728} & \num{1.81} & \num{1.74} & \num{1.78} & \num{1.12} & \num{1.80} \\
 \ch{Pb2BiSe2Cl3} & \cmcii & \num{4.22} & \num{13.46} & \num{17.34} & \num{-773} & \num{1.27} & \num{1.19} & \num{1.26} & \num{0.85} & \num{1.70} \\
 & \piic & \num{7.22} & \num{18.63} & \num{8.17} & \num{-786} & \num{2.36} & \num{2.19} & \num{2.36} & \num{2.48} & \num{1.97} \\\hline
 & \cmcm & \num{4.28} & \num{14.34} & \num{17.61} & \num{-667} & \num{1.37} & \num{1.14} & \num{1.42} & \num{0.58} & \num{1.34} \\
 \ch{Pb2BiSe2Br3} & \cmcii & \num{4.29} & \num{14.4} & \num{17.59} & \num{-703} & \num{1.20} & \num{1.07} & \num{1.26} & \num{0.45} & \num{1.10} \\
 & \piic & \num{7.23} & \num{16.96} & \num{8.50} & \num{-722} & \num{1.96} & \num{1.84} & \num{1.95} & \num{2.63} & \num{0.89} \\\hline
 & \cmcm & \num{4.38} & \num{14.15} & \num{16.64} & \num{-572} & \num{1.17} & \num{1.09} & \num{1.26} & \num{0.50} & \num{1.17} \\
 \ch{Pb2BiSe2I3} & \cmcii & \num{4.40} & \num{15.46} & \num{17.69} & \num{-600} & \num{1.07} & -/- & \num{1.26} & \num{0.38} & \num{0.82} \\
 & \piic & \num{7.45} & \num{17.04} & \num{8.76} & \num{-637} & \num{1.71} & -/- & \num{2.04} & \num{2.01} & \num{0.66} \\\hline
 & \cmcm & \num{4.33} & \num{12.46} & \num{17.78} & \num{-640} & \num{0.63} & -/- & \num{1.17} & \num{0.83} & \num{2.28} \\
 \ch{Pb2BiTe2Cl3} & \cmcii & \num{4.29} & \num{12.96} & \num{18.93} & \num{-629} & \num{0.40} & \num{0.39} & \num{0.45} & \num{0.99} & \num{0.98} \\
 & \piic & \num{6.36} & \num{18.47} & \num{8.45} & \num{-682} & \num{1.09} & \num{1.09} & \num{1.19} & \num{0.43} & \num{0.88} \\\hline
 & \cmcm & \num{4.34} & \num{12.65} & \num{18.28} & \num{-523} & \num{0.92} & -/- & \num{1.12} & \num{0.89} & \num{2.50} \\
 \ch{Pb2BiTe2Br3} & \cmcii & \num{4.30} & \num{13.40} & \num{19.06} & \num{-561} & \num{0.78} & \num{0.69} & \num{0.87} & \num{1.32} & \num{1.19} \\
 & \piic & \num{6.51} & \num{19.17} & \num{8.61} & \num{-604} & \num{1.21} & \num{1.12} & \num{1.15} & \num{2.21} & \num{1.21} \\\hline
 & \cmcm & \num{4.46} & \num{13.94} & \num{18.75} & \num{-419} & \num{0.74} & \num{0.54} & \num{0.83} & \num{0.63} & \num{0.83} \\
 \ch{Pb2BiTe2I3} & \cmcii & \num{4.48} & \num{15.3} & \num{18.79} & \num{-465} & \num{0.55} & \num{0.44} & \num{0.53} & \num{2.52} & \num{1.05} \\
 & \piic & \num{7.23} & \num{18.83} & \num{8.90} & \num{-494} & \num{0.98} & \num{0.50} & \num{1.05} & \num{0.84} & \num{1.38} \\\hline
\end{tabular}}
\vspace*{.17cm}
\footnoterule\footnotesize
\vspace*{-.5cm}
\begin{flushleft}
\textsuperscript{\hyperref[fn:tab_a_top]{\textit{a}}}\hspace*{0.0545cm} Lat. para.  are taken from RT ($T = \SI{293}{\kelvin}$) XRD meas. using Mo$K\alpha$ radiation[24] and band gap are taken from UV/vis absorption meas.[16] \label{fn:tab_a_bottom}\\[.25cm]
\textsuperscript{\hyperref[fn:tab_b_top]{\textit{b}}}\hspace*{0.0545cm} Lat. para. are taken from low-temp. ($T = \SI{173}{\kelvin}$) XRD meas. using Mo$K\alpha$ radiation.[24] \label{fn:tab_b_bottom}\\[.25cm]
\textsuperscript{\hyperref[fn:tab_c_top]{\textit{c}}}\hspace*{0.0545cm} Lat. para.  are taken from RT ($T = \SI{293}{\kelvin}$) XRD meas. using Mo$K\alpha$ radiation and band gap are taken from UV/vis absorption meas.[25] \label{fn:tab_c_bottom} \\[0.25cm]
\textsuperscript{\hyperref[fn:tab_d_top]{\textit{d}}}\hspace*{0.0545cm} Lat. para. for \cmcm are taken from RT ($T = \SI{293}{\kelvin}$) XRD meas. using Mo$K\alpha$ radiation[27] and band gap are taken photoconductivity meas.[88] \label{fn:tab_d_bottom} \\[0.25cm]
\textsuperscript{\hyperref[fn:tab_e_top]{\textit{e}}}\hspace*{0.0545cm} Lat. para. for \piic are taken from low-temp. ($T = \SI{100}{\kelvin}$) XRD meas. using Mo$K\alpha$ radiation.[27] \label{fn:tab_e_bottom} \\[0.25cm]
\textsuperscript{\hyperref[fn:tab_f_top]{\textit{f}}}\hspace*{0.0545cm} Lat. para.  are taken from RT ($T = \SI{293}{\kelvin}$) XRD meas. using Mo$K\alpha$ radiation and band gap are taken from UV/vis absorption meas.[25] \label{fn:tab_f_bottom}
\end{flushleft}
\end{minipage}
\end{table}
\clearpage

\clearpage

\pdfbookmark[section]{Random forest regression results for predicting the formation energies in \protect\cmcii and \protect\piic}{s6}
\section*{SM6: Random forest regression results for predicting the formation energies in \protect\cmcii and \protect\piic}\label{sec:si:s6}

\begin{figure}[h!]
\centering
\includegraphics[scale=0.95]{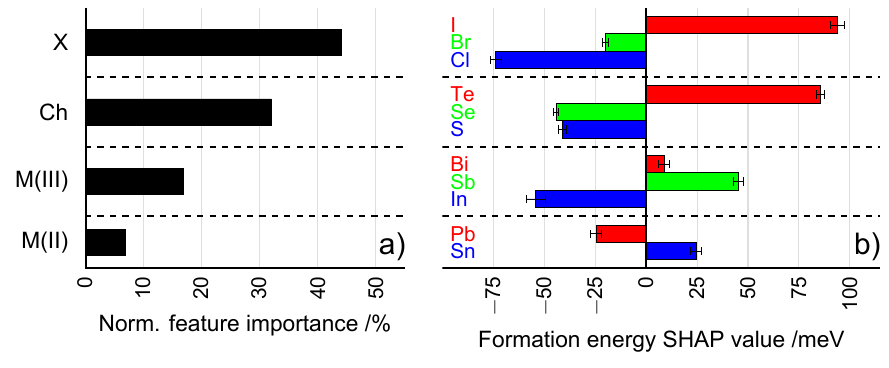}
\caption{Summary of the feature importance for the formation energy RF model in \protect\cmcii \pimmch compounds. a) Normalised atom site importance on the predicted formation energies in percent. b) Impact of elements within the \ch{M(II)}-, \ch{M(III)}-, \ch{Ch}- and \ch{X}-sites measured as mean shapley additive explanations (SAHP) values on the predicted base formation energy value in \unit{\milli\electronvolt}. The standard deviation for each mean SHAP value is displayed as error bars.}
\label{si:fig:formenergy_cmc21}
\end{figure}

\begin{figure}[h!]
\centering
\includegraphics[scale=0.95]{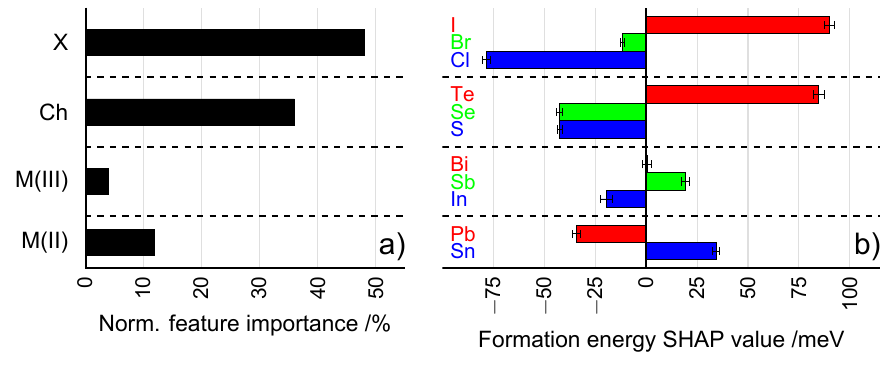}
\caption{Summary of the feature importance for the formation energy RF model in \protect\piic \pimmch compounds. a) Normalised atom site importance on the predicted formation energies in percent. b) Impact of elements within the \ch{M(II)}-, \ch{M(III)}-, \ch{Ch}- and \ch{X}-sites measured as mean SAHP values on the predicted base formation energy value in \unit{\milli\electronvolt}. The standard deviation for each mean SHAP value is displayed as error bars.}
\label{si:fig:formenergy_p21c}
\end{figure}

Note that the formation energy in Figures \ref{si:fig:formenergy_cmc21} and \ref{si:fig:formenergy_p21c} is negative, meaning that negative mean SHAP values indicate an increase in formation energy, whereas positive mean SHAP values indicate an decrease.

\clearpage

\pdfbookmark[section]{Random forest regression results for predicting the fundamental band gaps in \protect\cmcii and \protect\piic}{s7}
\section*{SM7: Random forest regression results for predicting the fundamental band gaps in \protect\cmcii and \protect\piic}\label{sec:si:s7}

\begin{figure}[h!]
\centering
\includegraphics[scale=1]{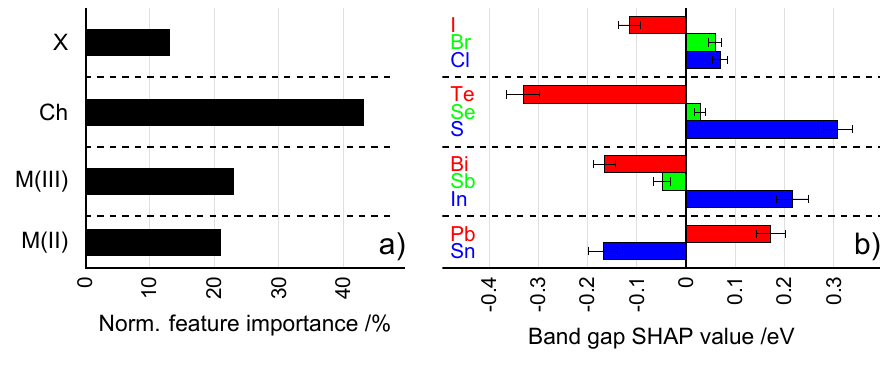}
\caption{Summary of the feature importance for the fundamental band gap RF model in \protect\cmcii \pimmch compounds. a) Normalised atom site importance on the predicted fundamental band gaps in percent. b) Impact of elements within the \ch{M(II)}-, \ch{M(III)}-, \ch{Ch}- and \ch{X}-sites measured as mean SAHP values on the predicted base fundamental band gap value in \unit{\electronvolt}. The standard deviation for each mean SHAP value is displayed as error bars.}
\label{si:fig:bandgap_cmc21}
\end{figure}

\begin{figure}[h!]
\centering
\includegraphics[scale=1]{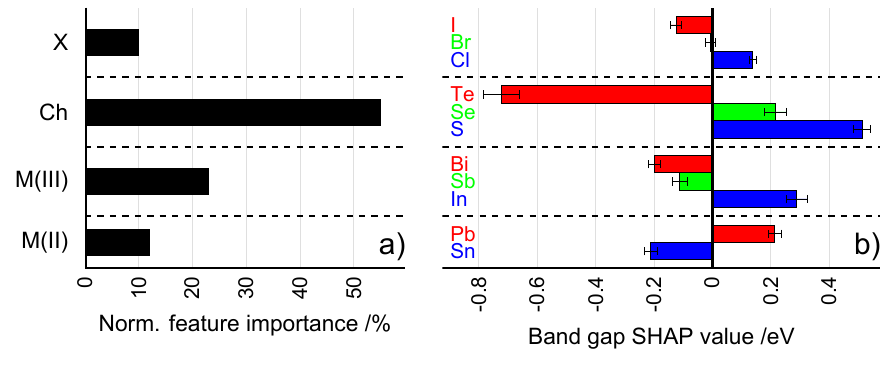}
\caption{Summary of the feature importance for the fundamental band gap RF model in \protect\piic \pimmch compounds. a) Normalised atom site importance on the predicted fundamental band gaps in percent. b) Impact of elements within the \ch{M(II)}-, \ch{M(III)}-, \ch{Ch}- and \ch{X}-sites measured as mean SAHP values on the predicted base fundamental band gap value in \unit{\electronvolt}. The standard deviation for each mean SHAP value is displayed as error bars.}
\label{si:fig:bandgap_p21c}
\end{figure}

\clearpage

\pdfbookmark[section]{Random forest regression results for predicting the lowest direct band gaps in \protect\cmcm, \protect\cmcii and \protect\piic}{s8}
\section*{SM8: Random forest regression results for predicting the lowest direct band gaps in \protect\cmcm, \protect\cmcii and \protect\piic}\label{sec:si:s8}

\begin{figure}[h!]
\centering
\includegraphics[scale=1]{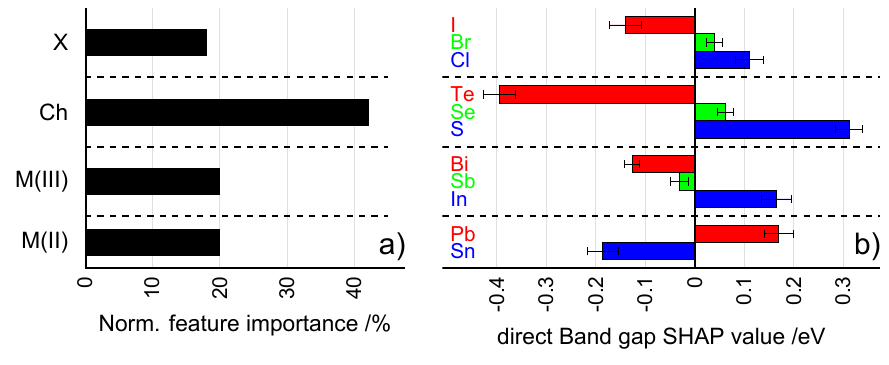}
\caption{Summary of the feature importance for the lowest direct band gap RF model in \protect\cmcm \pimmch compounds. a) Normalised atom site importance on the predicted lowest direct band gaps in percent. b) Impact of elements within the \ch{M(II)}-, \ch{M(III)}-, \ch{Ch}- and \ch{X}-sites measured as mean SAHP values on the predicted base lowest direct band gap value in \unit{\electronvolt}. The standard deviation for each mean SHAP value is displayed as error bars.}
\label{si:fig:bandgap_cmcm:direct}
\end{figure}

\begin{figure}[h!]
\centering
\includegraphics[scale=1]{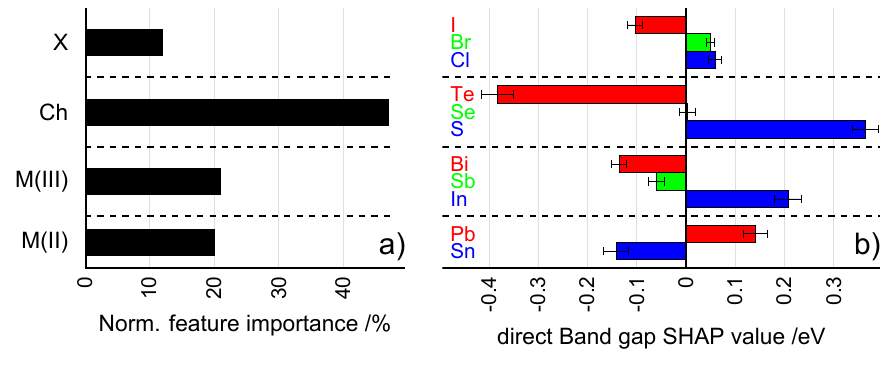}
\caption{Summary of the feature importance for the lowest direct band gap RF model in \protect\cmcii \pimmch compounds. a) Normalised atom site importance on the predicted lowest direct band gaps in percent. b) Impact of elements within the \ch{M(II)}-, \ch{M(III)}-, \ch{Ch}- and \ch{X}-sites measured as mean SAHP values on the predicted base lowest direct band gap value in \unit{\electronvolt}. The standard deviation for each mean SHAP value is displayed as error bars.}
\label{si:fig:bandgap_cmc21:direct}
\end{figure}

\begin{figure}[h!]
\centering
\includegraphics[scale=1]{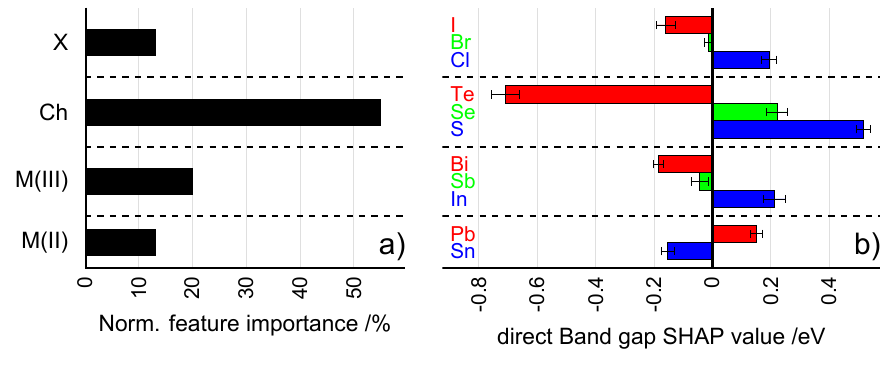}
\caption{Summary of the feature importance for the lowest direct band gap RF model in \protect\piic \pimmch compounds. a) Normalised atom site importance on the predicted lowest direct band gaps in percent. b) Impact of elements within the \ch{M(II)}-, \ch{M(III)}-, \ch{Ch}- and \ch{X}-sites measured as mean SAHP values on the predicted base lowest direct band gap value in \unit{\electronvolt}. The standard deviation for each mean SHAP value is displayed as error bars.}
\label{si:fig:bandgap_p21c:direct}
\end{figure}

In analogy to the discussion of the fundamental band gap in the main text [see Section 3.2], we determined the direct band gaps of all Perovskite-inspired quaternary mixed-metal chalcohalides (\pimmchs) per phase. In the case that the fundamental band gap is not a direct one, the lowest direct band gap between valence and conduction band has been determined for those indirect \pimmch compounds. In contrast to indirect band gaps, direct band gaps do not require an additional phonon to perform the transition. However, these direct gaps require higher photon energies than the minimal indirect band gaps within the material in order to be overtaken. Nevertheless, such transitions are also of interest e.g. for photovoltaic applications, so we calculated them as well and the results are shown in the following SI Figures \ref{si:fig:bandgap_cmcm:direct}, \ref{si:fig:bandgap_cmc21:direct} and \ref{si:fig:bandgap_p21c:direct} for \cmcm, \cmcii and \piic. Generally, we obtain analogous trends compared to the discussed fundamental band gaps in the main text, see Section 3.2 within the main text.

\clearpage

\pdfbookmark[section]{Random forest regression results for predicting the optical band gaps in \protect\cmcm, \protect\cmcii and \protect\piic}{s9}
\section*{SM9: Random forest regression results for predicting the optical band gaps in \protect\cmcm, \protect\cmcii and \protect\piic}\label{sec:si:s9}

Besides the fundamental band gaps [see main text Figure \ref{fig:results:bandgap}, as well as SI Figures in Sections \hyperlink{sec:si:s7}{S7} and \hyperlink{sec:si:s8}{S8}] we determined the optical band gap by means of a Tauc plot based on the adsorption spectra of the respective \pimmch material. Generally, there are two main findings for the optical band gaps. First, the optical band gaps are larger than the fundamental gaps, see SI table \ref{si:tab:matparameters}. This is caused by i) a low density of states at the band edges resulting from a low electronic degeneracy due to the low crystal symmetry (for \cmcm and \cmcii) as well as ii) a weak transition dipole moment between the valence and the conduction band minimum due to the symmetry restrictions and a low spatial overlap.[17] Kavanagh \textit{et al.} identified this mismatch between fundamental and optical band gap for \ch{Sn2SbS2I3}[17] and we confirmed this in our previous study fo]r lead-free \pimmchs.[20] Now, we observe this mismatch between fundamental and optical band gap for all 27 lead-free and 27 lead-based investigated \pimmchs per phase, see discussion in Section 3.2 main text. This indicates that such a mismatch occurs for all compounds within the \pimmch material space. Furthermore, we obtained analogue trends regarding the effect of atom sites and the elements on the optical band gap compared to the fundamental band gap, see discussion in Section 3.2 main text.

\begin{figure}[h!]
\centering
\includegraphics[scale=1]{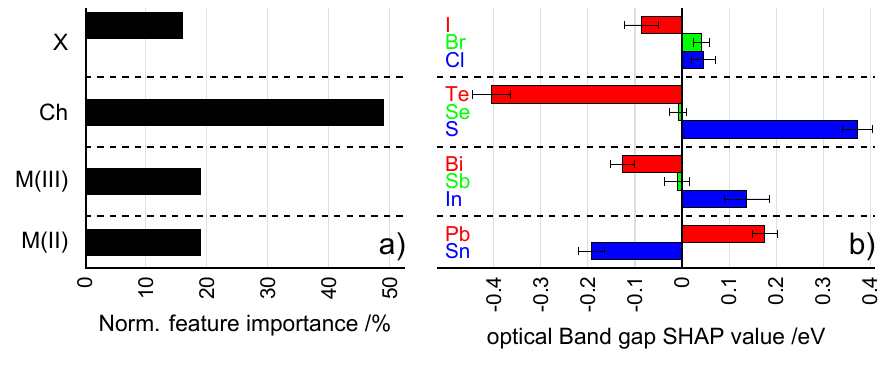}
\caption{Summary of the feature importance for the optical band gap RF model in \protect\cmcm \pimmch compounds. a) Normalised atom site importance on the predicted optical band gaps in percent. b) Impact of elements within the \ch{M(II)}-, \ch{M(III)}-, \ch{Ch}- and \ch{X}-sites measured as mean SAHP values on the predicted base optical band gap value in \unit{\electronvolt}. The standard deviation for each mean SHAP value is displayed as error bars.}
\label{si:fig:bandgap_cmcm:optical}
\end{figure}

\begin{figure}[h!]
\centering
\includegraphics[scale=1]{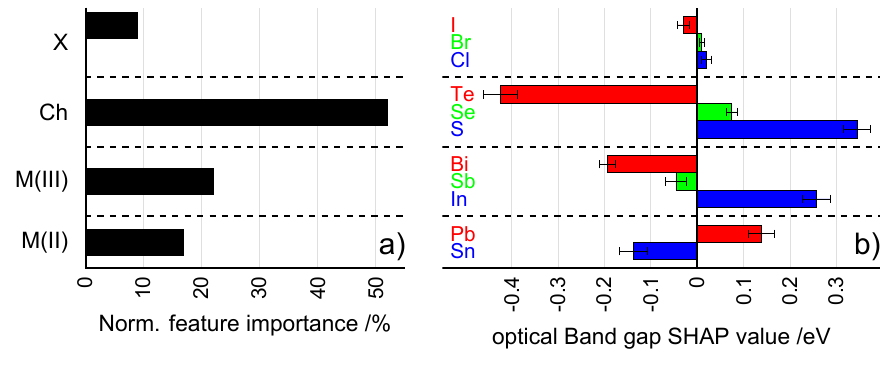}
\caption{Summary of the feature importance for the optical band gap RF model in \protect\cmcii \pimmch compounds. a) Normalised atom site importance on the predicted optical band gaps in percent. b) Impact of elements within the \ch{M(II)}-, \ch{M(III)}-, \ch{Ch}- and \ch{X}-sites measured as mean SAHP values on the predicted base optical band gap value in \unit{\electronvolt}. The standard deviation for each mean SHAP value is displayed as error bars.}
\label{si:fig:bandgap_cmc21:optical}
\end{figure}

\begin{figure}[h!]
\centering
\includegraphics[scale=1]{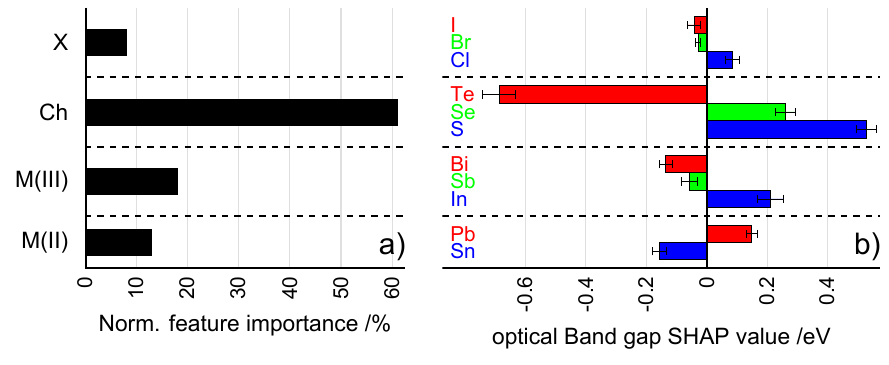}
\caption{Summary of the feature importance for the optical band gap RF model in \protect\piic \pimmch compounds. a) Normalised atom site importance on the predicted optical band gaps in percent. b) Impact of elements within the \ch{M(II)}-, \ch{M(III)}-, \ch{Ch}- and \ch{X}-sites measured as mean SAHP values on the predicted base optical band gap value in \unit{\electronvolt}. The standard deviation for each mean SHAP value is displayed as error bars.}
\label{si:fig:bandgap_p21c:optical}
\end{figure}

\clearpage

\pdfbookmark[section]{Random forest regression results for predicting the effective electron and hole masses in \protect\cmcii and \protect\piic}{s10}
\section*{SM10: Random forest regression results for predicting the effective electron and hole masses in \protect\cmcii and \protect\piic}\label{sec:si:s10}

\begin{figure}[h!]
\centering
\includegraphics[scale=1]{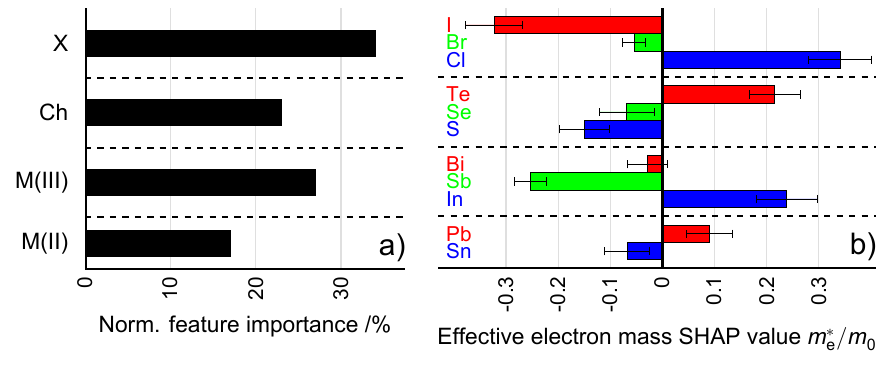}
\caption{Summary of the feature importance for the effective electron mass RF model in \protect\cmcii \pimmch compounds. a) Normalised atom site importance on the predicted effective electron masses in percent. b) Impact of elements within the \ch{M(II)}-, \ch{M(III)}-, \ch{Ch}- and \ch{X}-sites measured as mean SAHP values on the predicted base effective electron mass value per $m_0$ electron mass (\SI{9.109e-31}{\kilo\gram}). The standard deviation for each mean SHAP value is displayed as error bars.}
\label{si:fig:effmass_electron_cmc21}
\end{figure}

\begin{figure}[h!]
\centering
\includegraphics[scale=1]{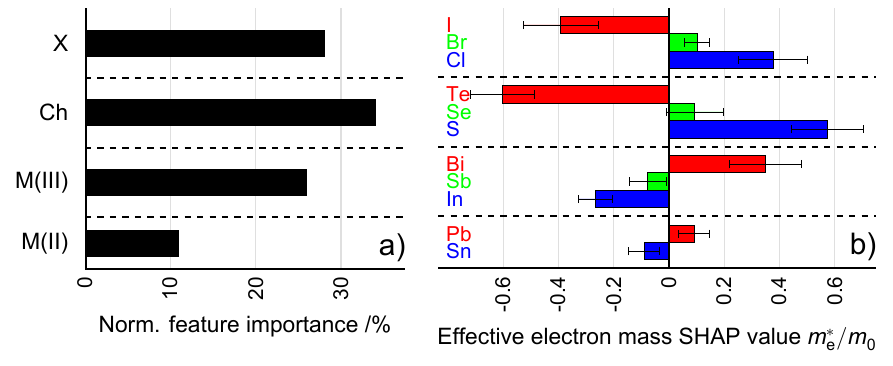}
\caption{Summary of the feature importance for the effective electron mass RF model in \protect\piic \pimmch compounds. a) Normalised atom site importance on the predicted effective electron masses in percent. b) Impact of elements within the \ch{M(II)}-, \ch{M(III)}-, \ch{Ch}- and \ch{X}-sites measured as mean SAHP values on the predicted base effective electron mass value per $m_0$ electron mass (\SI{9.109e-31}{\kilo\gram}). The standard deviation for each mean SHAP value is displayed as error bars.}
\label{si:fig:effmass_electron_p21c}
\end{figure}

\begin{figure}[h!]
\centering
\includegraphics[scale=1]{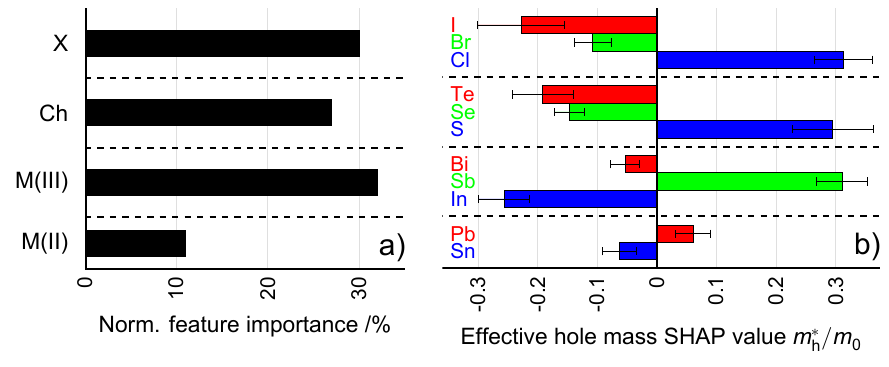}
\caption{Summary of the feature importance for the effective hole mass RF model in \protect\cmcii \pimmch compounds. a) Normalised atom site importance on the predicted effective hole masses in percent. b) Impact of elements within the \ch{M(II)}-, \ch{M(III)}-, \ch{Ch}- and \ch{X}-sites measured as mean SAHP values on the predicted base effective hole mass value per $m_0$ electron mass (\SI{9.109e-31}{\kilo\gram}). The standard deviation for each mean SHAP value is displayed as error bars.}
\label{si:fig:effmass_hole_cmc21}
\end{figure}

\begin{figure}[h!]
\centering
\includegraphics[scale=1]{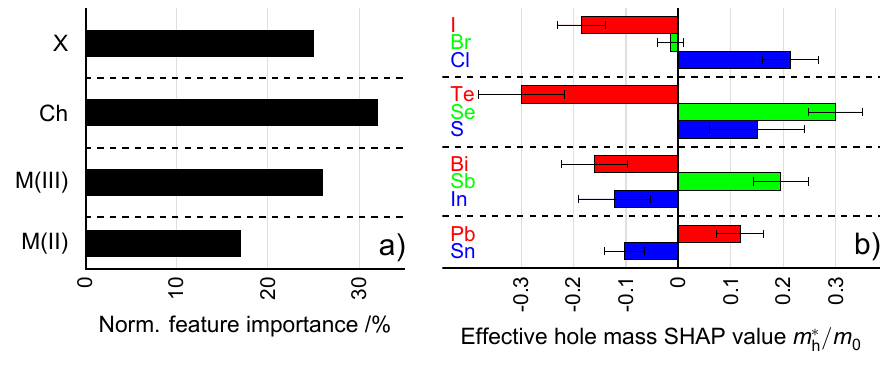}
\caption{Summary of the feature importance for the effective hole mass RF model in \protect\piic \pimmch compounds. a) Normalised atom site importance on the predicted effective hole masses in percent. b) Impact of elements within the \ch{M(II)}-, \ch{M(III)}-, \ch{Ch}- and \ch{X}-sites measured as mean SAHP values on the predicted base effective hole mass value per $m_0$ electron mass (\SI{9.109e-31}{\kilo\gram}). The standard deviation for each mean SHAP value is displayed as error bars.}
\label{si:fig:effmass_hole_p21c}
\end{figure}

\clearpage

\pdfbookmark[section]{Density of states of lead-free and lead-based \protect\ch{M(II)2BiSe2Br3} compounds in \protect\cmcii and \protect\piic phases}{s11}
\section*{SM11: Density of states of lead-free and lead-based \protect\ch{M(II)2BiSe2Br3} compounds in \protect\cmcii and \protect\piic phases}\label{sec:si:s11}

\begin{figure}[h!]
\centering
\includegraphics[scale=0.9]{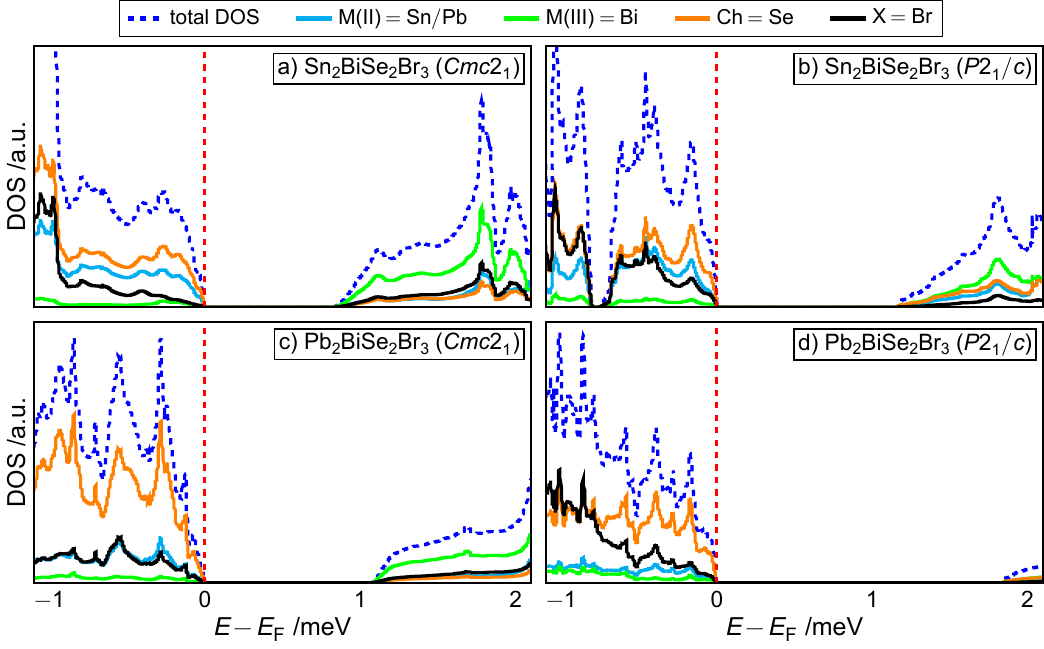}
\caption{Density of state for \protect\ch{Sn2BiSe2Br3} at a) \protect\cmcii and b) \protect\piic phases as well as for \protect\ch{Pb2BiSe2Br3} at c) \protect\cmcii and \protect\piic phases.}
\label{si:fig:dos}
\end{figure}

\clearpage

\pdfbookmark[section]{Band structures of lead-free and lead-based \protect\ch{M(II)2BiSe2Br3} compounds in \protect\cmcm, \protect\cmcii and \protect\piic phases}{s12}
\section*{SM12: Band structures of lead-free and lead-based \protect\ch{M(II)2BiSe2Br3} compounds in \protect\cmcm, \protect\cmcii and \protect\piic phases}\label{sec:si:s12}

\begin{figure}[h!]
\centering
\includegraphics[width=0.925\linewidth]{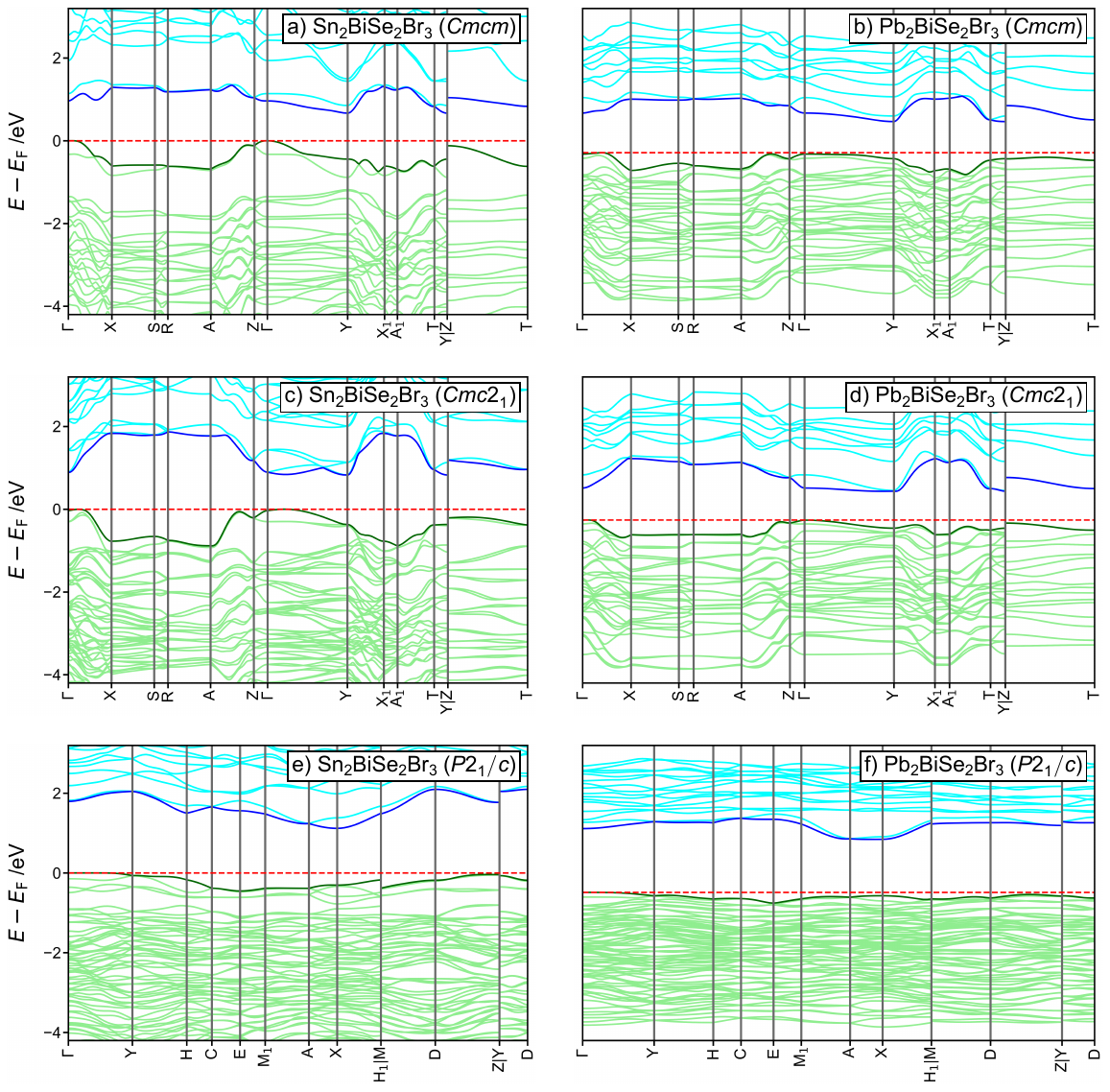}
\caption{Band structures for \protect\ch{Sn2BiSe2Br3} at a) \protect\cmcm, c) \protect\cmcii and e) \protect\piic phases as well as for  \protect\ch{Pb2BiSe2Br3} at b) \protect\cmcm, d) \protect\cmcii and f) \protect\piic phases. Colors: Light green denotes the valence bands, with the highest occupied band highlighted in dark green. Cyan denotes the conduction bands, with the lowest unoccupied band highlighted in blue. }
\label{si:fig:bandstructure}
\end{figure}

\clearpage

\pdfbookmark[section]{Comparison of \protect\pimmchs material properties in \protect\cmcii and \protect\piic}{s13}
\section*{SM13: Comparison of \protect\pimmchs material properties in \protect\cmcii and \protect\piic}\label{sec:si:s13}

\begin{figure}[h!]
\centering
\includegraphics[scale=.65]{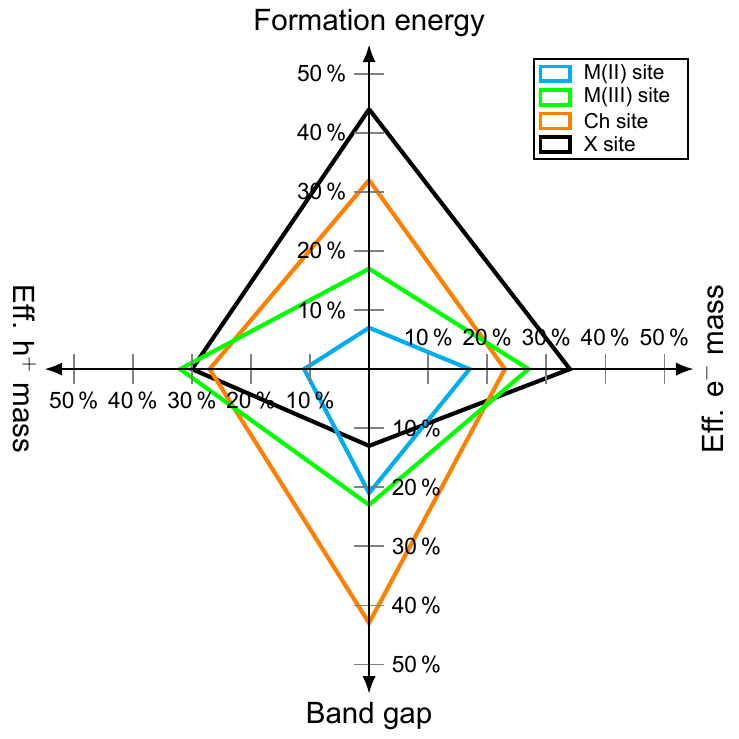}
\caption{Summary of atom site importance on the formation energy, band gap, as well as the effective masses of electrons and holes in \protect\cmcii.}
\label{si:fig:vgl_featureimpoetance_cmc21}
\end{figure}

\begin{figure}[h!]
\centering
\includegraphics[scale=.65]{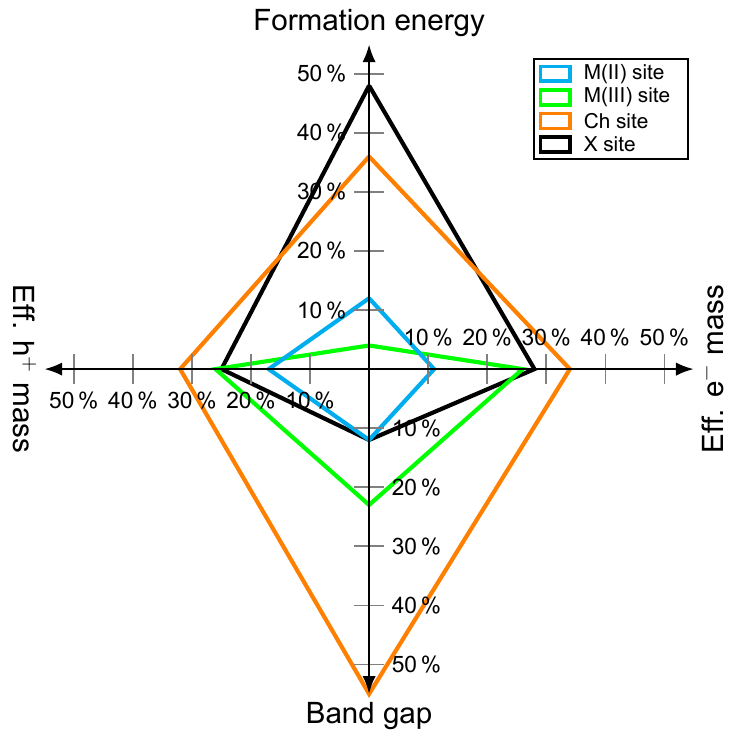}
\caption{Summary of atom site importance on the formation energy, band gap, as well as the effective masses of electrons and holes in \protect\piic.}
\label{si:fig:vgl_featureimpoetance_p21c}
\end{figure}